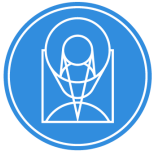

Instrument Science Report WFC3 2024-07

# Revisiting x-CTE in WFC3/UVIS

Jay Anderson
June 7, 2024

**ABSTRACT**

*Most discussion of charge-transfer-efficiency (CTE) losses involves the parallel transfer of charge in the y-direction down the chip columns (y-CTE). Serial charge-transfer efficiency (x-CTE) refers to the horizontal transer of charge. Serial CTE losses were first assessed in WFC3/UVIS in 2014 (WFC3/ISR 2014-02), where it was found that bright stars were shifted by ~0.0015 pixel away from the readout amplifier. Now that the WFC3/UVIS CCD has spent more than three times as long in space, imperfect serial CTE should have three times the impact on the images, so we revisit the effect, characterizing serial CTE in a similar manner to our models for parallel-CTE by using a combination of warm pixels, cosmic rays, saturated stars, and overscan pixels. Overall, serial CTE has a much smaller impact on images than does parallel CTE. As of 2023, parallel CTE can have a ~5% impact on bright sources and a >50% impact on faint sources, and serial CTE can have a ~1% impact on bright sources and a ~3% impact on fainter sources. The first few pixels in the serial CTE trails are much sharper than the parallel trails, but there is a very faint component to the serial trails that extends much farther than the parallel trails — even wrapping around to the next row. We develop a pixel-based model for the trapping and release of charge in the serial register and release a stand-alone beta-version of this pixel-based serial-CTE correction. Most images are not significantly impacted by the x-CTE effect, however HST users that require high-precision astrometry could benefit from this correction — at the very least, so that they can quantify its impact on their science.*

## 1. Introduction

Serial CTE losses in WFC3/UVIS were last studied in 2014, when WFC3/UVIS had only been on board HST for 5 years (Anderson, 2014). At that time, bright stars far from the serial register were found to be shifted by 0.0015 pixel away from the register, while fainter stars were shifted by 0.004 pixel. At the time, these small shifts were not relevant for many science programs, but it was useful to be able to quantify the amplitude of the effect, if only to assure the community that it was not yet a major source of concern. Now in 2024, WFC3/UVIS has been in the harsh radiation environment of low-earth orbit for about three times as long as it had in 2014. On the downside, this extra time in space both makes serial CTE more significant to science, but on the upside, the more significant amplitude makes it easier to characterize.

In this ISR, we will do a deep dive into serial CTE in an effort to understand how it impacts the pixels we download from the detector. Serial CTE shares a lot of commonalities with the better-studied parallel CTE (for the latest review, see Anderson *et al*. 2021). They both appear to arise from charge traps that delay the shuffling of some charge from one pixel to another. The result of this delay is a loss of charge in bright pixels and a release of charge into fainter "upstream" pixels. There are also some differences between serial and parallel CTE, related to the way the serial register is used to read out rows of pixels during the readout process.

We begin this study in **Section 2** by looking at an anomaly that was discovered in a Quicklook[1] image of a bias frame. This anomaly exhibited what appeared to be an extremely energetic cosmic-ray event that had a CTE trail that wrapped around from the end of one row to the beginning of the next row. In an attempt to explain this phenomenon, **Section 3** goes into the details of the WFC3/UVIS readout mechanics.

**Section 4** then uses the same strategy that has been used to study parallel CTE to examine serial CTE, namely by examining the trails behind warm/hot pixels in the dark exposures. This allows us to study serial CTE for faint and medium-bright stimulus pixels. But unfortunately (for this study), WFC3/UVIS has very few bright hot pixels so the dark exposures cannot help us characterize serial CTE for pixels with more than 25,000 electrons. So, in **Section 5** we extend our analysis to the horizontal virtual overscan region, which allows a very precise study of the impact of imperfect serial CTE on both small and extremely large charge packets.

**Section 6** assembles all of the above results into a model of trapping and release that is similar to the model used to describe parallel CTE. **Section 7** incorporates the forward model into a FORTRAN subroutine that is able predict the blurring induced by serial CTE transfer. Since serial-CTE blurring is very much still at the perturbation level (~1%), we find that one iteration is sufficient to use the forward model to predict the original pixel distribution from an observed image. We evaluate this correction and in **Section 8** provide it in several formats to the community.

---

[1] See Medina *et al*. (2021) for a description of WFC3's Quicklook procedure.

## 2. Strange features seen in bias

The original impetus for this study was a curious feature noticed in a Quicklook[2] survey of some bias images. **Figure 1** shows the bias image, along with some horizontal averages.

Note that the feature exhibits not only a trail to the right – as would be expected from CTE in the serial register (aka, x-CTE) — but there is also a faint trail in the same rows that can be seen emanating from the *left* of the detector and appearing to fade out just before the CR. Furthermore, this trail is present not only in the on-sky pixels, but it is also present in the 30 horizontal overscan pixels on the right, and in the 25 physical pre-scan pixels on the left. The trail maintains an amplitude of 1 electron for 300 pixels beyond the CR, and it maintains an average amplitude of more than 0.1 electron for almost 2000 serial shifts. This document is a deep-dive effort to understand, characterize, and provide a correction for this phenomenon.

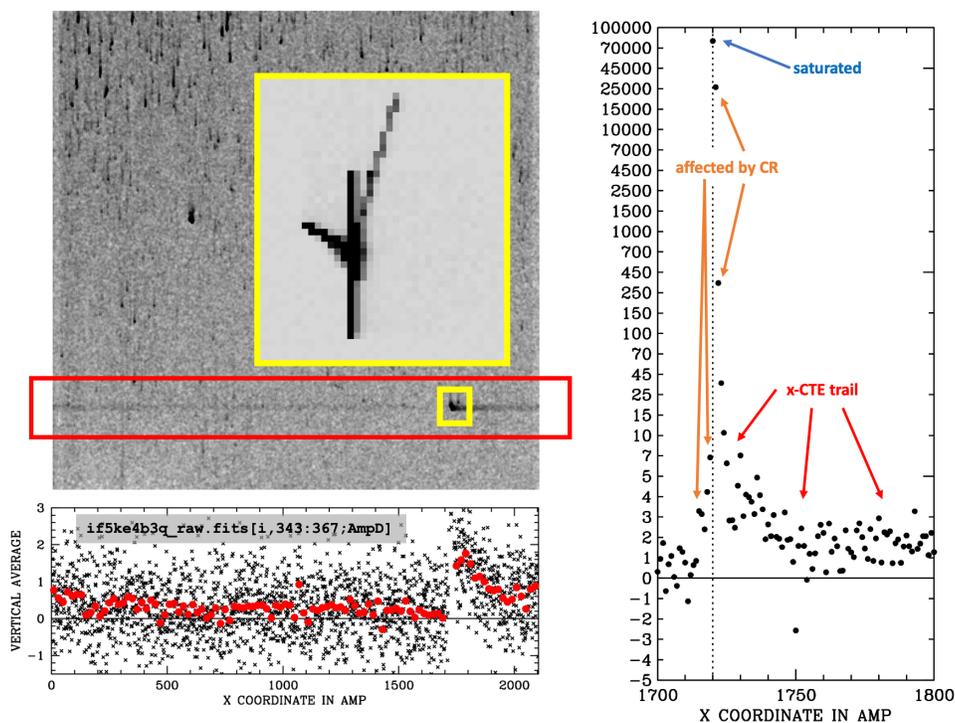

**Figure 1: Block averaged representation of image `if5ke4b3q`, an unflashed bias image. Note the increase in CRs with distance from the bottom of the image due to CR hits that arrive during the 90 seconds of readout. The CR under study is shown in the yellow outline. The bottom graph shows the average flux per column in the 24 rows affected by the CR. Red dots correspond to averages over ten columns. The right graph shows the immediate vicinity of the CR. The image shown is 2103×2070 and includes 25 physical pre-scan pixels on the left, 30 horizontal overscan pixels on the right and 19 vertical overscan pixels at the top.**

---

[2] See Medina et al. (2021). The WFC3 team inspects every image that is downloaded from the telescope via the "quicklook" interface, which allows various features to be characterized and documented, such as dragon's breath, satellite trails, internal reflections, etc.

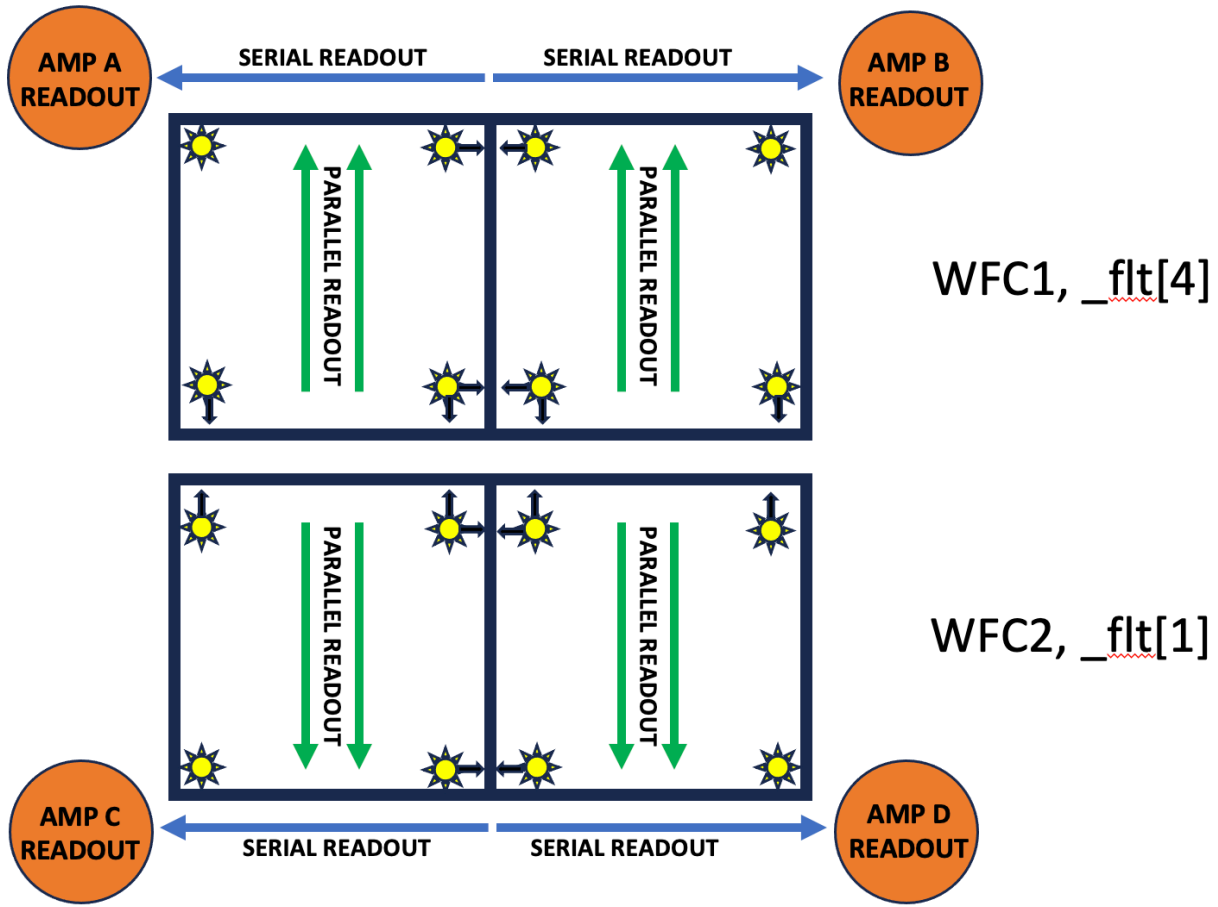

**Figure 2:** A schematic showing the WFC3/UVIS detector readout for the two CCD chips. The top chip is WFC1 and is in the first extension of `_flt` images; the bottom chip is WFC2 and is in the first extension of `_flt` images. Stars close to the amplifiers show minimal signs of CTE losses, but stars far from the amplifier in the parallel direction can find themselves shifted away from the amplifier by CTE in the parallel direction (y-CTE) in the direction of the chip-gap. Stars far from the amplifier in the serial direction find themselves shifted away from the amplifier in the serial direction (x-CTE) in the direction of the boundary between amplifiers.

### 3. Background on WFC3/UVIS readout mechanics

**Figure 2** provides a schematic of the WFC3/UVIS detector and how the two CCD chips are read out. There are four amplifiers, and each chip is read out by two amplifiers. Amplifiers A and B read out UVIS1 (the top chip) and amplifiers C and D read out UVIS2 (the bottom chip). Physical overscan columns run along the outer left and right edges of the field of view while horizontal virtual overscan columns reside between amps A/B and C/D. Vertical virtual overscan rows (not used for this study) fall along the chip edges near the chip gap. See the Instrument Handbook (Marinelli & Dressel 2024) for more details.

Since this document will be dealing with both parallel- and serial-transfer issues, we will convert the `_raw` images into a convenient format that we will call the `_raz` format. In these images, basic bias subtraction and de-trending has been performed and the 2103×2070 pixels that correspond to each quadrant are oriented such that the parallel shifts are downward and the serial shifts are to the left.

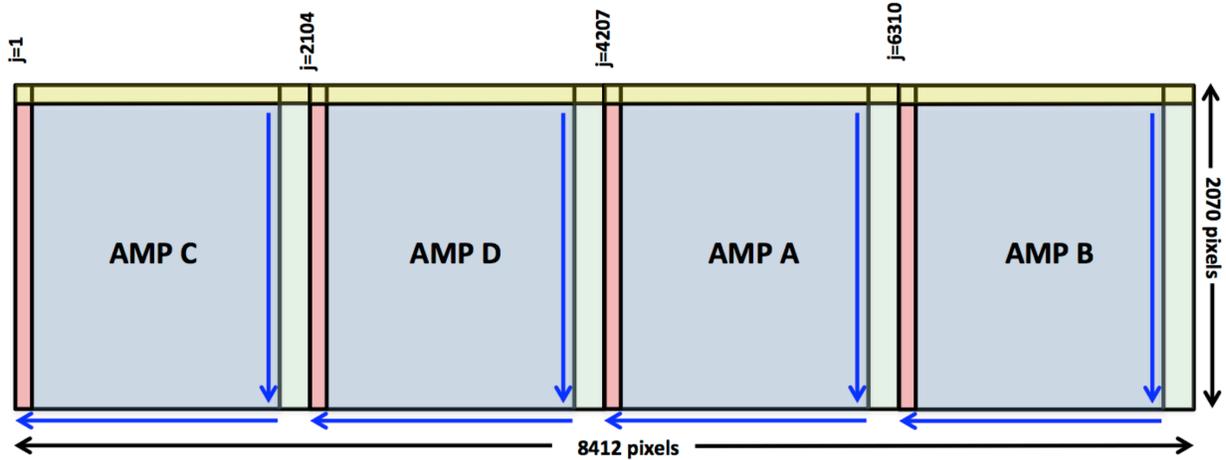

**Figure 3:** Setup of the wide "raz" image format (reproduced from ISR/WFC3 2014-22, Anderson & Baggett). The light blue regions are the 2048×2051 on-sky pixels in each amp. The red regions correspond to the 25×2071 horizontal physical pre-scan pixels. The green regions correspond to the 30×2071 horizontal virtual overscan pixels. Finally, the yellow regions along the top correspond to the vertical virtual overscan pixels (not used for this study). Readout directions are marked with blue arrows.

**Figure 3** shows a schematic of the `_raz` format images. The first 25 columns in each amplifier block correspond to physical pre-scan pixels. They are physical pixels, but they are not exposed to the sky. The next 2048 pixels (columns 0026-2073) are on-sky pixels. Finally, the final 30 columns, 2074 to 2103 are horizontal virtual overscan pixels. The horizontal virtual overscan pixels do not correspond to real, physical pixels, but rather they correspond to what ends up in the serial register pixels when the electrons from the real pixels are shifted out of them. In **Section 5**, we will delve deeper into the readout mechanics.

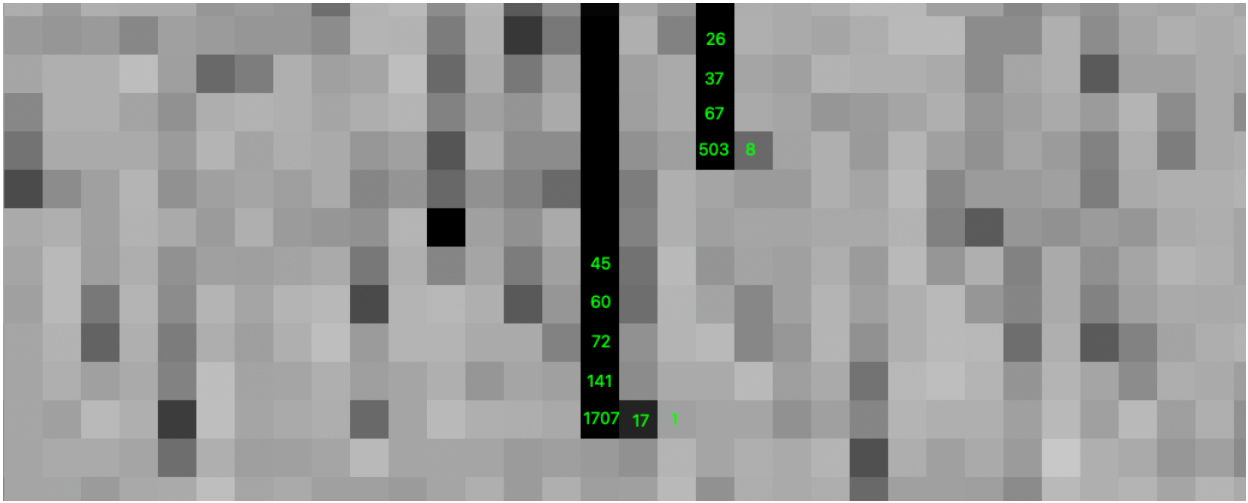

Figure 4: Image cutout near the upper right corner of quadrant C (i.e., furthest from the C amplifier), where both x- and y-CTE losses and trailing should be maximal. The pixel values associated with two hot pixels are identified.

## 4. Examining Warm Pixels in the Darks

In their initial models of parallel charge-transfer for the ACS detector, Anderson & Bedin (2010) analyzed dark exposures to characterize the behavior of the trails upward from warm pixels. The amount of charge in the trails corresponded to the number of electrons lost from the warm pixels.

## 4.1 The Cal-program darks

We stacked ~1 month worth of 900s dark images taken with a post-flash of 20 e⁻ (between Jan 8, 2024 and Feb 7, 2024). There were 75 darks, and all were taken in program CAL-17347 (PI – Pidgeon). The image in **Figure 4** shows a 32x12 region of that stack in the upper right corner of quadrant C where both parallel and serial CTE losses would be near maximum. The local background has been subtracted so that the pixel values directly reflect the impact of the hot/warm pixels on their environment.

Two hot pixels have been identified in the image and the values of their pixels have been labeled. It is clear that the vertical upward trails associated with imperfect parallel y-CTE are much more significant than the horizontal rightward trails associated with serial x-CTE. The first pixel in the vertical trail is about ten times the first pixel in the horizontal trail. Also, the vertical trail has a much more gradual slope than the steep horizontal trail. Nevertheless, the horizontal trail contains about 1% of the flux of the hot pixel in both cases. A shift of 1% of the flux by 1 pixel could introduce an astrometric shift on the order of 1% of a pixel (modulo PSF effects). As such, imperfect serial CTE can certainly have a measurable effect on our measurements, given that bright stars in images can typically be measured with a precision of better than 1% of a pixel (Anderson 2022).

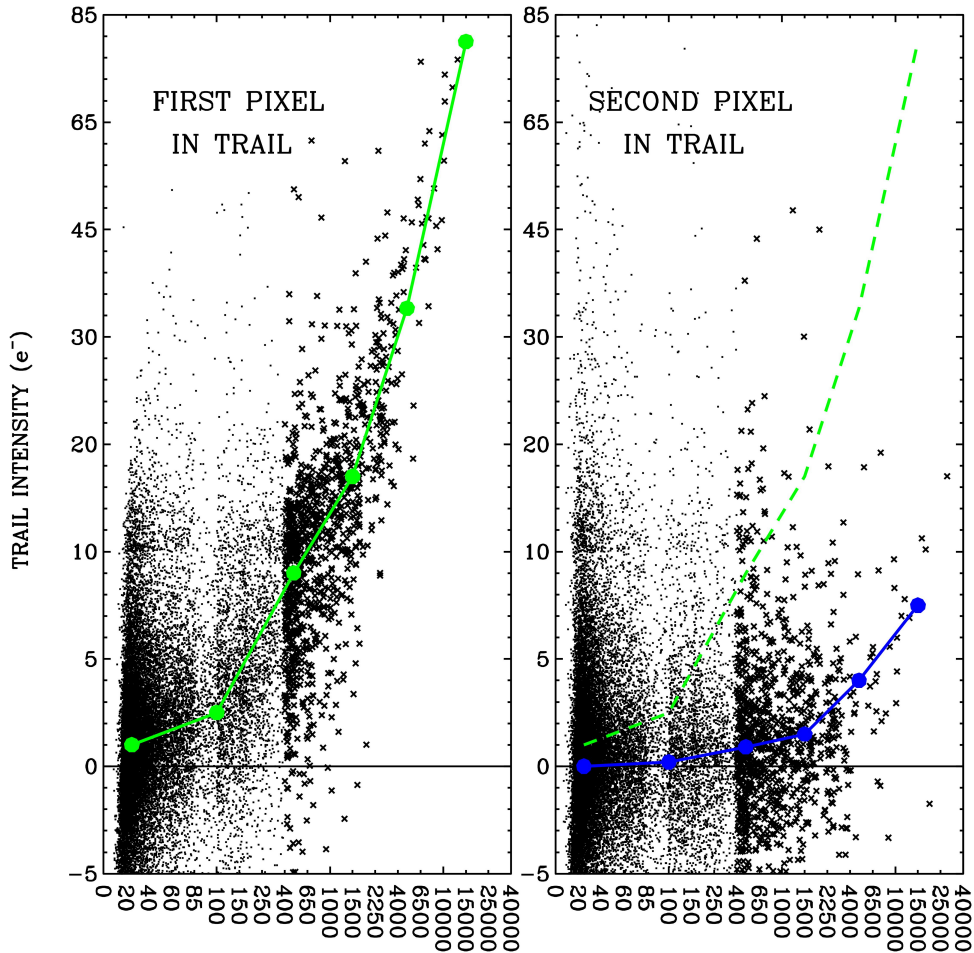

**Figure 5:** (Left) The number of electrons over the local background in the first pixel to the right of hot/warm pixels in the early 2024 post-flashed 900s darks, as a function of the hot/warm pixel intensity. The green line traces the general trend. (Right) The intensity of the second pixel as a function of hot/warm pixel intensity.

We measured all hot/warm pixels in the dark stack that produced a local maximum in at least 50 out of the 75 exposures. We examined the vertical and horizontal trails extending upward and to the right of the hot/warm pixels.

**Figure 5** shows the first pixel in the serial-CTE trail as a function of the intensity of the stimulating pixel for hot/warm pixels between 1750 and 2000 serial shifts from the register (where x-CTE losses would be expected to be largest). Note that even after 15 years on orbit, there are still not very many hot pixels with 900-second intensities greater than 25,000 electrons. This makes it difficult to measure the average first pixel intensity for extremely bright pixels. Also, since the background in these dark images is post-flashed to 20 electrons, the lowest bin (at 25 electrons) is not particularly reliable. We will clearly have to use other datasets to measure the trails behind the bright warm pixels and behind fainter ones. Finally, the right panel shows that the intensity in the second trail pixel is down by about a factor of ten relative to the first trail pixel, a much sharper drop than is seen in parallel CTE (Anderson 2020).

Table 1: The general trend of the first and second pixels in the serial and parallel trails from the early 2024 dark analysis.

| WP/HP | Serial CTE (x) | | Parallel CTE (y) | |
|---|---|---|---|---|
| | First Pixel | Second Pixel | First Pixel | Second Pixel |
| 25 e$^-$ | 1.0 e$^-$ (4%) | 0.0 e$^-$ | 13 e$^-$ (52%) | 7 e$^-$ |
| 100 e$^-$ | 2.5 e$^-$ (2.5%) | 0.2 e$^-$ | 30 e$^-$ (30%) | 14 e$^-$ |
| 500 e$^-$ | 9 e$^-$ (1.8%) | 0.9 e$^-$ | 65 e$^-$ (13%) | 30 e$^-$ |
| 1500 e$^-$ | 17 e$^-$ (1.3%) | 1.5 e$^-$ | 110 e$^-$ (7%) | 50 e$^-$ |
| 5000 e$^-$ | 34 e$^-$ (0.7%) | 4.0 e$^-$ | 175 e$^-$ (3.5%) | 90 e$^-$ |
| 15000 e$^-$ | 80 e$^-$ (0.3%) | 7.5 e$^-$ | 350 e$^-$ (2.3%) | 160 e$^-$ |

**Table 1** provides the trends shown in **Figure 5** in table form, along with similar measurements for parallel (y-)CTE. Note again that the ratio between the first trail pixel in the parallel and serial directions is about 10% for moderately faint WPs (100 e$^-$). But it is closer to 25% for bright hot pixels (15,000 e$^-$), i.e. the serial and parallel trails have a different spectrum of traps. The parallel CTE traps have a much steeper spectrum, meaning that fainter WPs are much more affected by parallel CTE than are brighter ones.

## 4.2 Is the effect due to x-CTE losses?

Thus far we have interpreted the rightward trailing as being CTE-related. But, in principle, it could be related to something else, such as some kind of hysteresis or bias-shift effect in the readout amplifier. To satisfy ourselves that this truly is CTE-related, we verify two aspects: (1) that the effect is roughly linear with the number of serial transfers (more transfers are likely to experience more traps), and (2) that the effect has been growing linearly over the time since WFC3/UVIS was installed and began its exposure to the harsh radiation environment of low earth orbit.

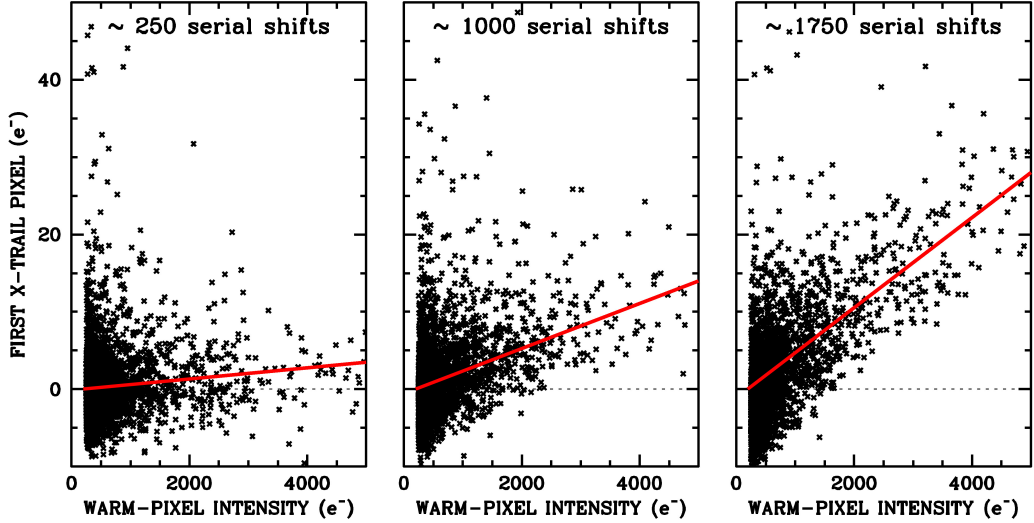

**Figure 6:** The trail intensity is proportional to the number of serial shifts, consistent with x-CTE. Left panel shows the first trail pixel for WPs that suffer very few serial shifts. Middle panel shows the trails from the middle of the serial register. Right panel shows trails from WPs that are shifted across almost the entire 2000 pixels of the serial register.

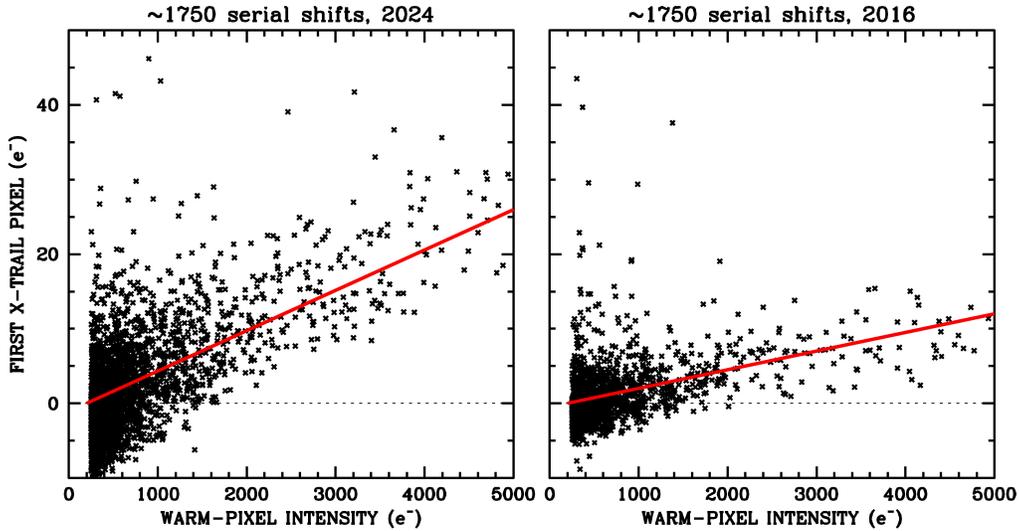

**Figure 7:** The trail intensity increases linearly with time, consistent with x-CTE. Left: the first pixel in the x-trail for WPs far from the serial readout in 900-s darks from 2024. Right: the same selection for 900-s darks taken in 2016, when the detector had been in space for about half as long as in 2024.

**Figure 6** shows that the trails that we have been examining do indeed have the expected behavior with column number, $i$, namely that the trail intensity is proportional to the number of serial shifts undergone by the warm pixel. The same behavior is seen for parallel CTE.

**Figure 7** shows that the signals in the trails do indeed increase linearly with time. The panel on the left shows the first pixel in the x-trail for darks taken in early 2024, and the right panel shows the same quantity for darks taken in late 2016. The trail signals have clearly increased by a factor of about two as the detector has aged by a factor of two (installation was in mid 2009), as expected from a radiation-damage-induced phenomenon.

We conclude that since the trails are linear in both time and in the number of serial shifts, it is reasonable to attribute this phenomenon to serial CTE. In the remainder of this study, we will characterize serial CTE losses from the smallest electron packets to the largest and also pin down the full trail profile.

## 4.3 x-CTE at the low end: unflashed darks

Most darks are now taken with 20 electrons of postflash, which is the recommended minimum background level for users to avoid pathological parallel CTE losses (see Chapter 6, WFC3 Instrument Handbook). This image background allows the darks to be better matched to typical observations. A pair of unflashed darks is taken each month for calibration monitoring purposes. These darks can help us understand the serial losses for the small electron packets.

We identified 44 unflashed darks taken in the year between November 2022 and October 2023 and stacked them in a similar manner to the analysis above. We examined the serial CTE trails for the smaller packets and were able to construct the trends that are shown in **Figure 8** and provided in **Table 2**.

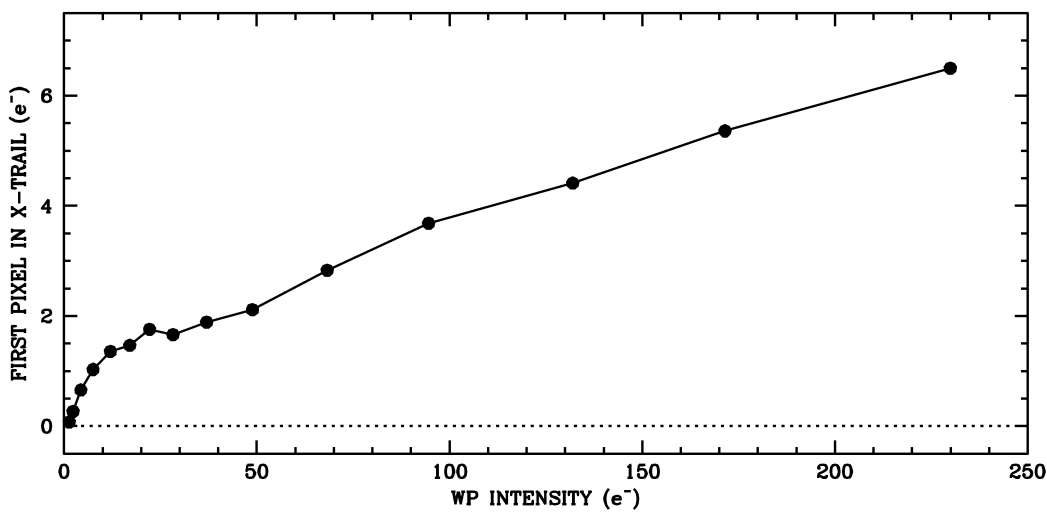

**Figure 8:** Trends of first serial trail pixel against warm pixel (WP) intensity from the unflashed darks taken between Nov22 and Oct 23.

**Table 2:** The trend shown in Figure 6 of the first serial trail pixel (STP#1) against WP intensity from the unflashed (low-background) darks.

| WP | STP#1 | WP | STP#1 | WP | STP#1 |
|---|---|---|---|---|---|
| 1 | 0.07 | 17 | 1.46 | 68 | 2.83 |
| 2 | 0.27 | 22 | 1.76 | 95 | 3.68 |
| 4 | 0.65 | 28 | 1.66 | 132 | 4.41 |
| 7 | 1.03 | 37 | 1.88 | 170 | 5.36 |
| 12 | 1.35 | 48 | 2.11 | 229 | 6.50 |

The shape of the WP-versus-trail curve is reminiscent of that seen for y-CTE, i.e. a large number of traps affect the first few electrons in a pixel (cf. Figure 7 of Anderson et al. 2021). **Figure 8** shows there is a steep rise in x-CTE from zero to about 20 electrons then a largely linear increase in the number of impacting traps as the WP increases from 20 electrons to 250 electrons. This behavior is quite different from that for parallel y-CTE, where the trap density rises roughly with the square-root of the WP intensity beyond the knee at 20 electrons.

## 4.4 The serial trail

The steepness and faintness of the serial x-CTE trails makes it difficult to study the trail profile much beyond the first trail pixel, even though the impetus for this study (the extreme cosmic ray event shown in **Figure 1**) shows that the serial trails can be thousands of pixels long.

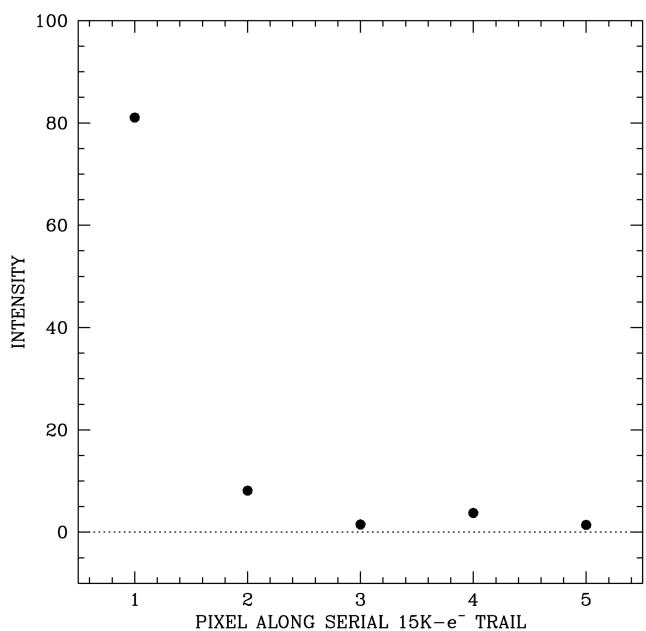

**Figure 9: Intensity of pixels along the serial trail following hot pixels with intensities of greater than 15,000 electrons.**

**Figure 9** shows the intensity of the first five pixels in the serial trail for all hot pixels with more than 15,000 electrons in them. The first pixel in the trail contains about 80 e⁻, the second pixel ~8 e⁻, the third ~2 e⁻, the fourth ~ 4 e⁻ and the fifth around 1 e⁻. One of the reasons this trail is hard to measure is that there are only *five* hot pixels in this brightness range that are more than 1750 transfers from the serial register. While there are many more lower intensity hot/warm pixels, given how faint the trails are, it will be harder to measure their extended trails.

To construct a reasonable trail profile and characterize the trap distribution that affects electron clouds with a decent fraction of full well (~60K e⁻), we need another source of high-intensity stimuli.

# 5  Finding more bright pixels to study

Since there are not enough hot pixels brighter than 15,000 electrons to allow us to probe the x-CTE properties all the way up to full-well, we considered other bright stimuli, such as bright stars and CRs. Bright stars have PSF halos, and it can be extremely difficult to remove these bright halos well enough to uncover the serial-CTE trail underneath, particularly as the highest signal-to-noise part of the x-CTE trail is the first pixel in the trail, adjacent to the stellar flux peak pixel. We then considered studying cosmic rays in the dark exposures. One benefit of CRs is that they tend not to be as extended as stars. CRs typically affect between 1 and 10 pixels, though extremely high-energy CRs can often affect dozens of pixels. The bright CR shown in **Figure 1** is quite anomalous, with a primary and a secondary trail, as well as a long bloom of pixels affected by the overload of charge that got deposited in the central pixel. Even though CRs are much tighter than stars, they still tend to affect multiple pixels, and this makes it hard to make a clean measurement of the flux in the first trail pixel (which is what we are using to baseline the strength of x-CTE).

## 5.1 The horizontal virtual overscan

Given that hot pixels, bright stars, and cosmic rays in the on-sky pixels are all unsuitable for characterizing the x-CTE signal in the trail, we investigated using the horizontal overscan pixels. The horizontal virtual overscan regions are denoted in light-green in **Figure 3**.

When studying y-CTE several years ago, we explored using the vertical virtual overscan pixels (seen in the light-yellow regions of **Figure 3**) to help pin down the pixel-based y-CTE model. We found, though, that the vertical virtual overscan region is not ideal for this because the parallel readout takes 90 seconds, and during this time the overscan pixels both get filled with dark current and get hit by cosmic rays. In addition, the three upper rows of science pixels have different physical sizes and thus intercept different amounts of light from astrophysical sources, sky, post-flash, or lamps. This creates a complicated stimulus for detailed modeling of the vertical overscan.

Thankfully, the horizontal overscan does not suffer from any of these deficiencies. The serial readout takes only 46.2 ms (see Table 6 from Baggett et al. 2003), so there is not much time for the pixels in the horizontal overscan to fill with dark current or get pocked by CRs. As such, these pixels should register zero absent the x-CTE charge trails and they should provide a relatively clean way to measure the x-CTE trails.

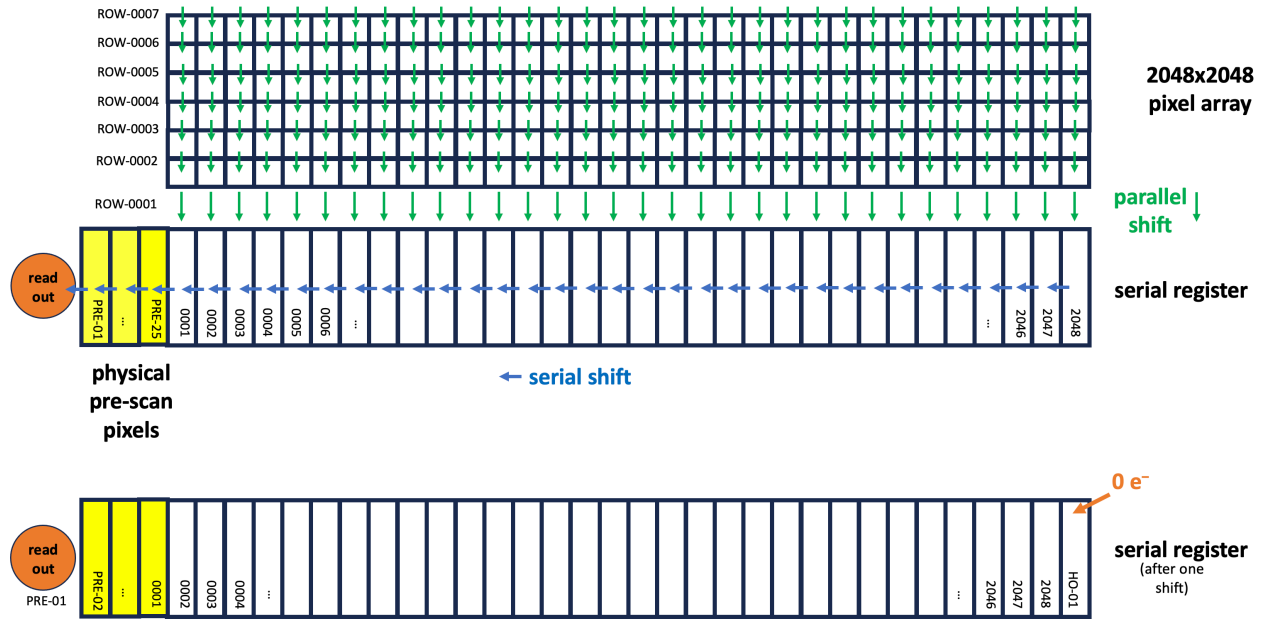

Figure 10: At top, schematic showing how the 2048×2051 on-sky pixels of each half-chip are parallel-shifted downwards into the serial register for read out. Note that the serial register is shown here with pixels that are intentionally "taller" than the regular pixels. The serial register is designed with deeper wells in order to handle larger charge packets than the ~70,000 e⁻ maximum of the regular pixels. These larger register pixels allow for binned-mode readouts that preserve the full dynamic range. The 25 yellow pixels on the left correspond to the physical pre-scan. At bottom, the schematic of the serial register shows its status after one serial shift to the left has occurred.

The blue arrows in **Figure 3** show how the 2048×2051 on-sky pixels of each amplifier are parallel shifted downwards, one row at a time, into the serial register. After each shift downward, the serial register is shifted leftward to the readout. **Figure 10** provides a more detailed schematic of this parallel-then-serial readout process for the first row of on-sky pixels. The green arrows show the 2048 on-sky pixels getting shifted down into the serial register. Once this takes place, they are shifted serially leftward (blue arrows) to the readout amplifier.

The top depiction of the serial register above shows what it looks like after the first row of on-sky pixels gets parallel shifted downward into the register. The readout of the register then proceeds. The serial register is then shifted to the left one pixel. The first pre-scan pixel gets shifted into the readout amplifier and read out. The state of the serial register after this first shift and read is shown in the bottom schematic of **Figure 10**.

Note that the rightmost pixel (2048) gets shuffled to the left by one pixel, but *nothing* gets shuffled into the place that pixel 2048 used to occupy in the serial register. The pixel that was to the right of that pixel in the on-sky image (pixel i=2049) got parallel-shifted downward into a different serial register (for the next amp over). We label this new end pixel HO-01 (for horizontal overscan pixel #1), and it starts out with zero electrons.

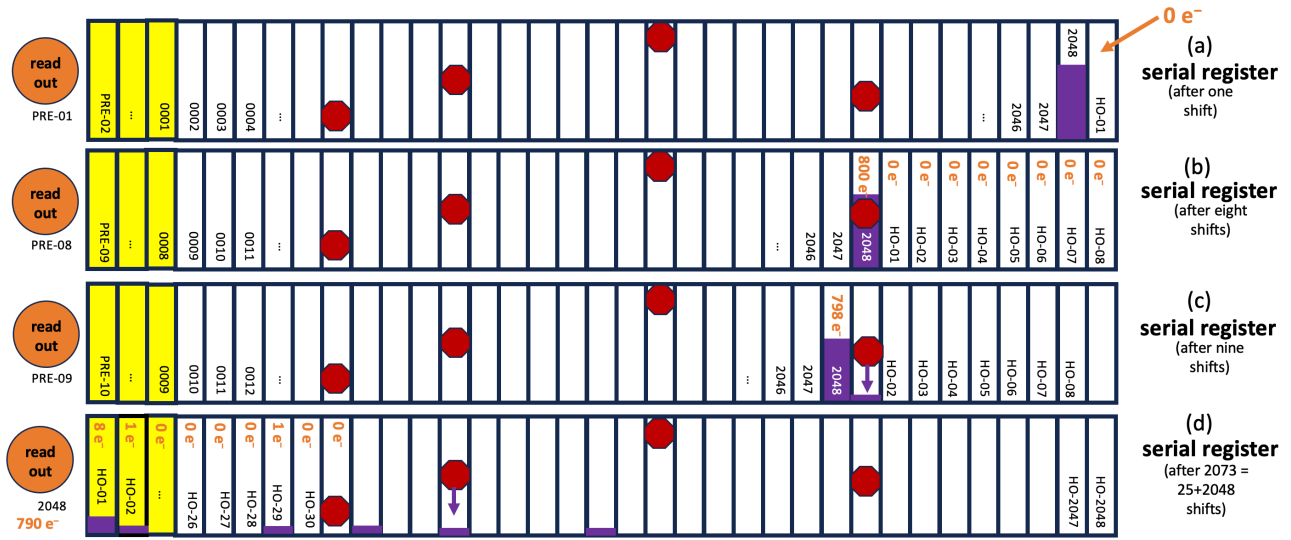

**Figure 11:** Schematic showing the serial register after various numbers of shifts to illustrate the charge trapping, trailing, and readout with 1 shift, 8 shifts, 9 shifts, and 2073 shifts in panels a, b, c, and d, respectively. The purple block signifies a moderate number of electrons in the last on-sky pixel in the row (pixel #2048) in panel (a). The red stop signs correspond to traps that remove some number of electrons from pixel clouds that are big enough to reach the traps.

**Figure 11** shows how the serial shifting proceeds in more detail. The top schematic (a) shows the serial register with the same timing as the bottom panel of the previous figure, but now with a moderately large charge packet in pixel #2048 of the row. Panel (b) shows the packet later encountering a trap that can grab an electron or two from a moderate-sized charge packet. Panel (c) shows the charge that survives the trap moving left one more pixel. The trap then releases one electron into the first pixel of the trail.

Panel (d) shows pixel the same bright pixel #2048 finally reaching the readout amplifier after 2073 serial shifts. The first horizontal overscan pixel is now in the first register pixel, ready to be read out next. This pixel, which is the first pixel in pixel #2048's trail, now has eight extra electrons in it, thanks to the charge released from various traps that pixel #2048 encountered along its multiple shifts through the serial register.

Since the horizontal overscan is initially populated with zeros and collects signal only from traps in the serial register that release electrons (the dark current during its rapid serial journey should be entirely negligible), it should be possible to study the x-CTE trails directly in the overscan pixels without interference from any astrophysical complications.

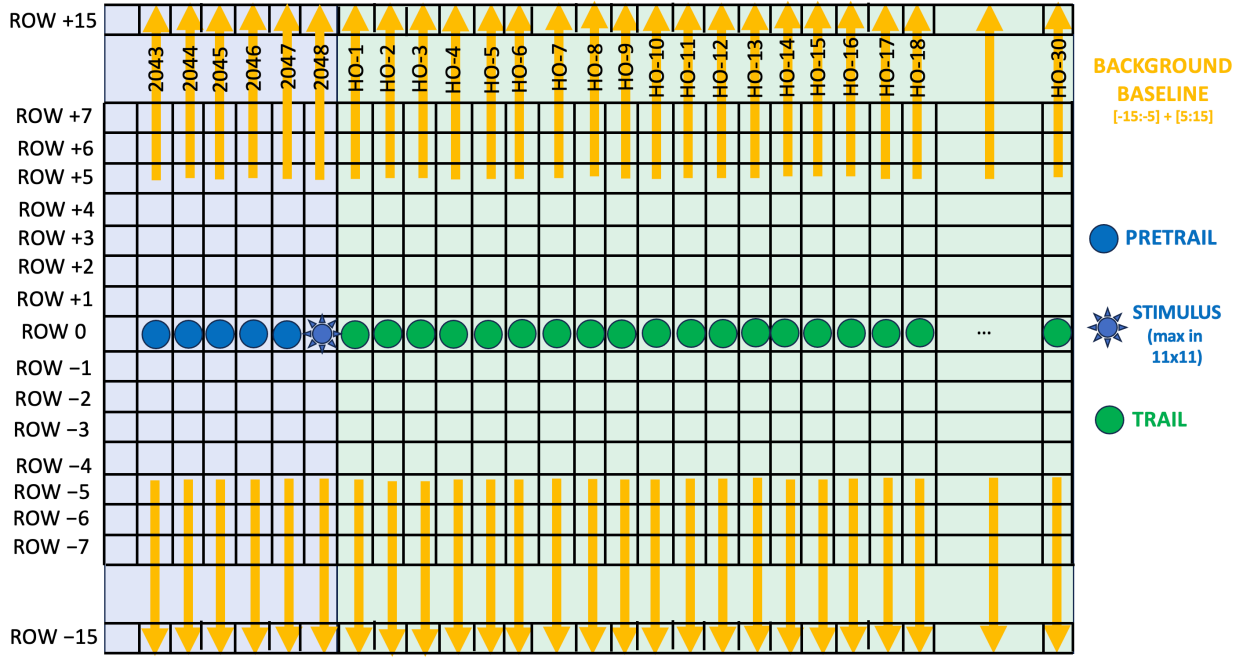

Figure 12: Schematic showing how we examined the trails in the virtual horizontal overscan. The blue-shaded area represents the on-sky pixels. The green-shaded area represents the virtual horizontal overscan. The "stimulus" pixel was the maximum in the 11x11 surrounding pixels. For each of the preceding and trail pixels, we subtracted the sigma-clipped background from the 32 pixels above and below, as shown.

## 5.2 Examining trails in the horizontal overscan

To identify useful overscan trails to study, we examined all the raw full-frame WFC3/UVIS GO images taken within the calendar year 2023 (4642 images). For each image we performed bias-subtraction and converted the _raw images into the _raz format shown in **Figure 3**. We then evaluated every pixel in the 2048[th] on-sky column and identified those that were the brightest pixels within an 11×11-pixel box, finding 642,869 such pixels. For each of these, we extracted the 5 pixels to the left, the target pixel [2048], and the 30 horizontal overscan pixels in the same row. In order to properly baseline the amount of charge, we also determined and subtracted the background of the 32 pixels that were between 15 and 30 pixels above and below each target pixel. The schematic in **Figure 12** shows the pixels under analysis.

Our focus is on target stimulus pixels with sky backgrounds between 10 and 40 electrons and background RMSs less than 20. **Figure 8** indicates that the serial CTE in the first trail pixel from the ~20 electrons of background should be on the order of 2 electrons. This signal has been removed by our background-baseline subtraction.

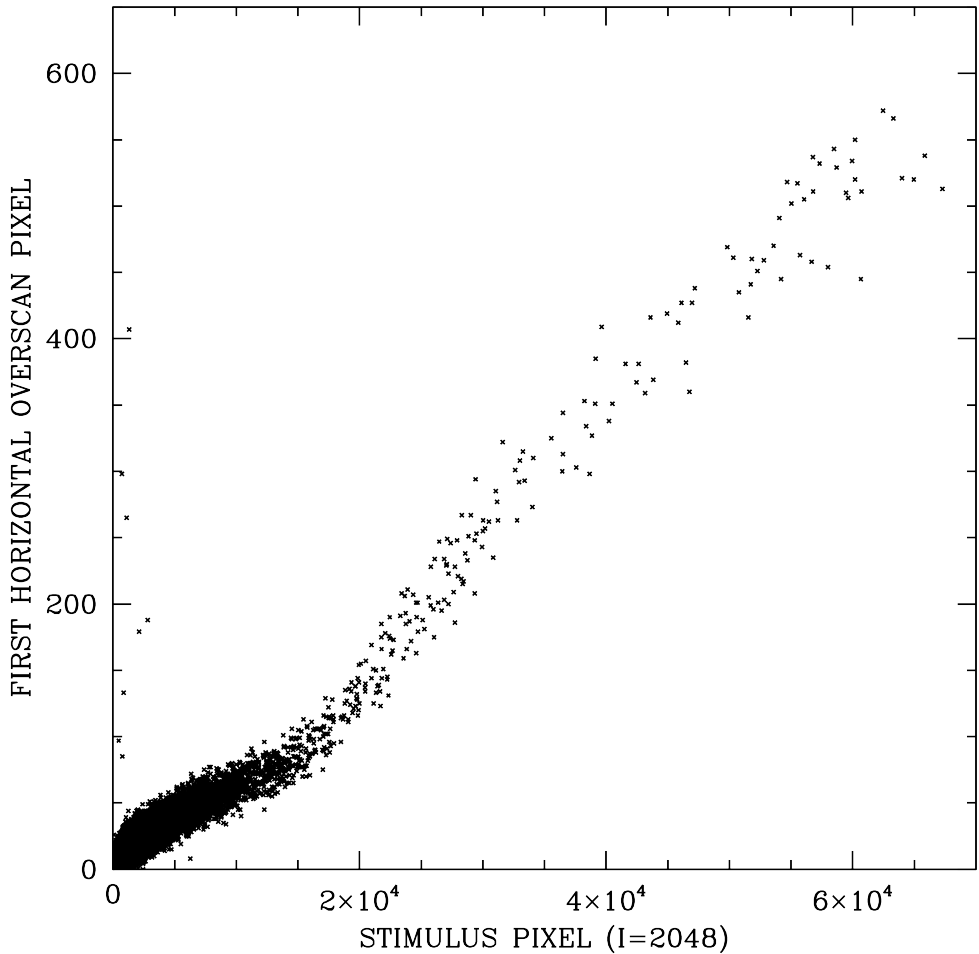

**Figure 13: The first pixel in the horizontal overscan as a function of the stimulus pixel (the last pixel in the on-sky pixels). Note the linear scale on both axes.**

In **Figure 13**, the flux in the baseline-subtracted first overscan pixel (HO-01) is plotted as a function of the stimulus pixel (i=2048, the last on-sky pixel). Recall that our analysis of the existing warm/hot pixels had allowed exploration of the x-CTE trails for pixels only as bright as 25,000 electrons. But now, by using the overscan pixels, we can explore the trails for pixel levels all the way up to full-well.

As **Figure 13** shows, pixels that have a full-well's worth of electrons (60,000 e⁻) have about 1% of those electrons (600 e⁻) in the first trail pixel. The plot has a linear scale on both axes, and to a first approximation, the 1% rule holds for a large fraction of stimulus levels, although there is a knee at about 15,000 electrons where the level in the first pixel of the trail is a bit lower.

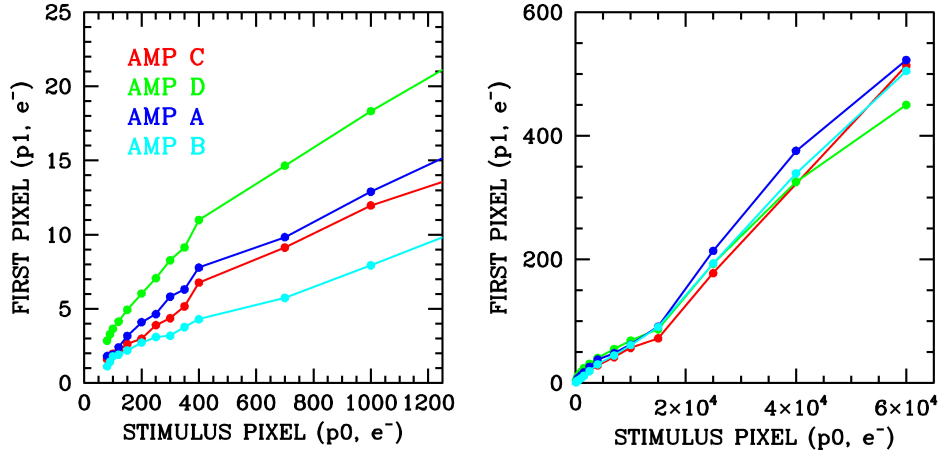

Figure 14: The flux in the first pixel of the horizontal overscan row as a function of the stimulus pixel for the four different amplifiers. At left is a zoom in to low stimulus values, and at right is the full range.

We binned the data from **Figure 13** by amplifier (chip quadrant) and plotted the results separately in **Figure 14**. The bins are non-overlapping, so each point is statistically independent of the other points. The fact that the four amplifiers show consistently different trends implies that the four have different trap distributions.

**Figures 13** and **14** summarized the trend of stimulus pixel against HO-1 (horizontal overscan pixel 1), the first pixel in the trail. We examine more trail pixels in **Figure 15**. The left panel shows the average trail in the 30 horizontal overscan pixels behind the brightest stimulus pixels. (The first pixel in the trail is off the plot at 390 e$^-$.) The initial part of the trail is clearly very steep while the rest of the trail declines gradually, down to 2 e$^-$ at the right edge of the plot.

The right panel of **Figure 15** illustrates how similar the trails are behind six different levels of stimulus pixels. All trails are scaled by the initial pixel in the trail (p1). In general, the trails exhibit similar sharp drops and gradual declines.

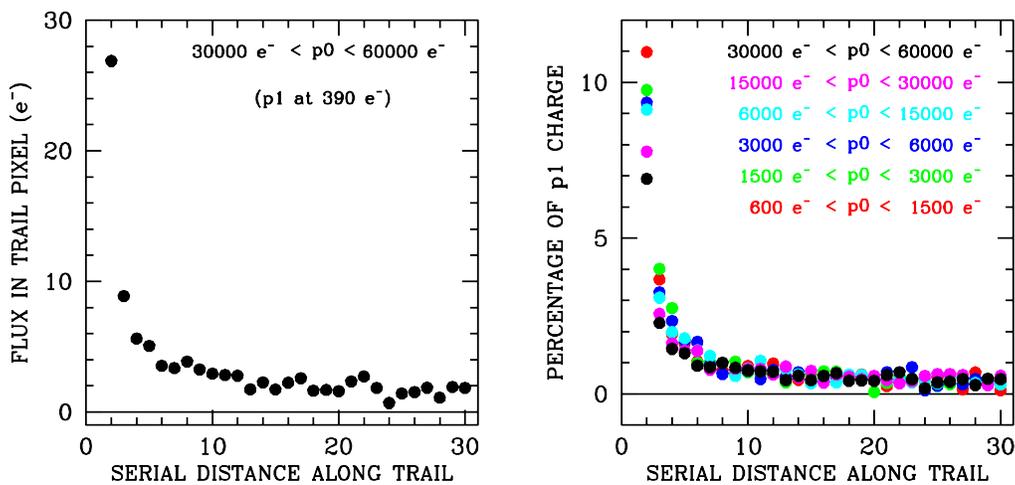

Figure 15: (Left) The full trail (except for the first pixel) in the virtual horizontal overscan for the brightest stimulus pixels. The trail declines rapidly then has a long component with measurable flux (2 e$^-$). (Right) The trails for variety of stimuli levels, scaled by the intensity of the first trail pixel.

## 5.3 Residual signal in the serial register

The initial impetus for this study was to investigate the very bright cosmic-ray event that was seen in a very low background image with a trail that extended in a way that wrapped around to the next row. We are now at a point where we can see how charge can get wrapped around into the next row.

The schematic in **Figure 16** continues the serial readout procedure where the schematic in **Figure 11** left off. During the serial readout process, the charge packets in the serial register pixels get shifted leftward to the readout amplifier. Nothing gets shifted into these pixels to replace the charge packets, so these "virtual" pixels begin their trip across the serial register empty.

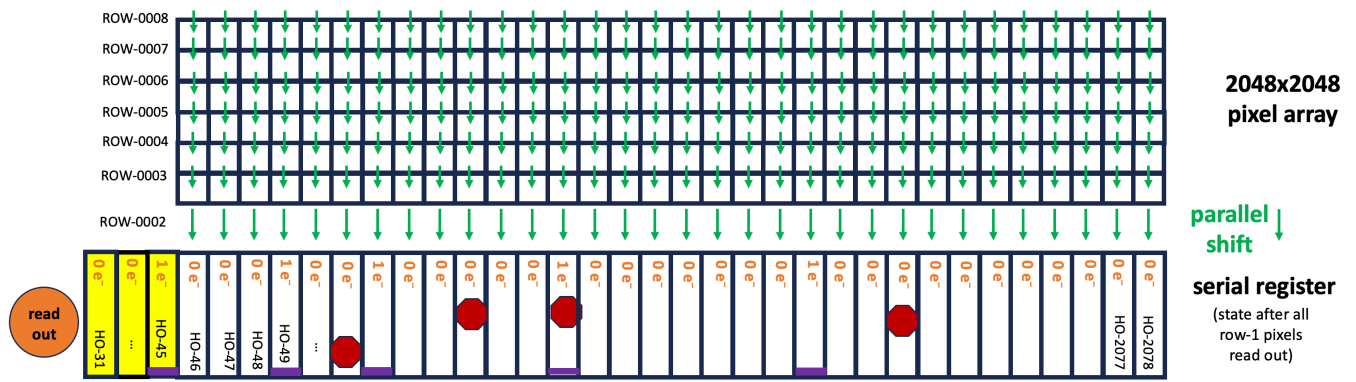

**Figure 16:** The state of the serial amplifier after the first row is read out (c.f. Fig 11) It is clear that the serial register is not necessarily empty when the next row is parallel-shuffled in.

But as these initially empty pixels are shuffled leftward towards the serial register, charge traps that had been filled by electron packets from the previously shuffled on-sky pixels ended up releasing their trapped electrons into these previously empty pixel packets. As such, these once-empty pixels can contain a dusting of charge. The register is in this "dusted" state when the next row of on-sky pixels gets parallel-shuffled down into the readout register, and the register will contain *both* contributions. This how charge trails of one row can end up contributing charge to the row above.

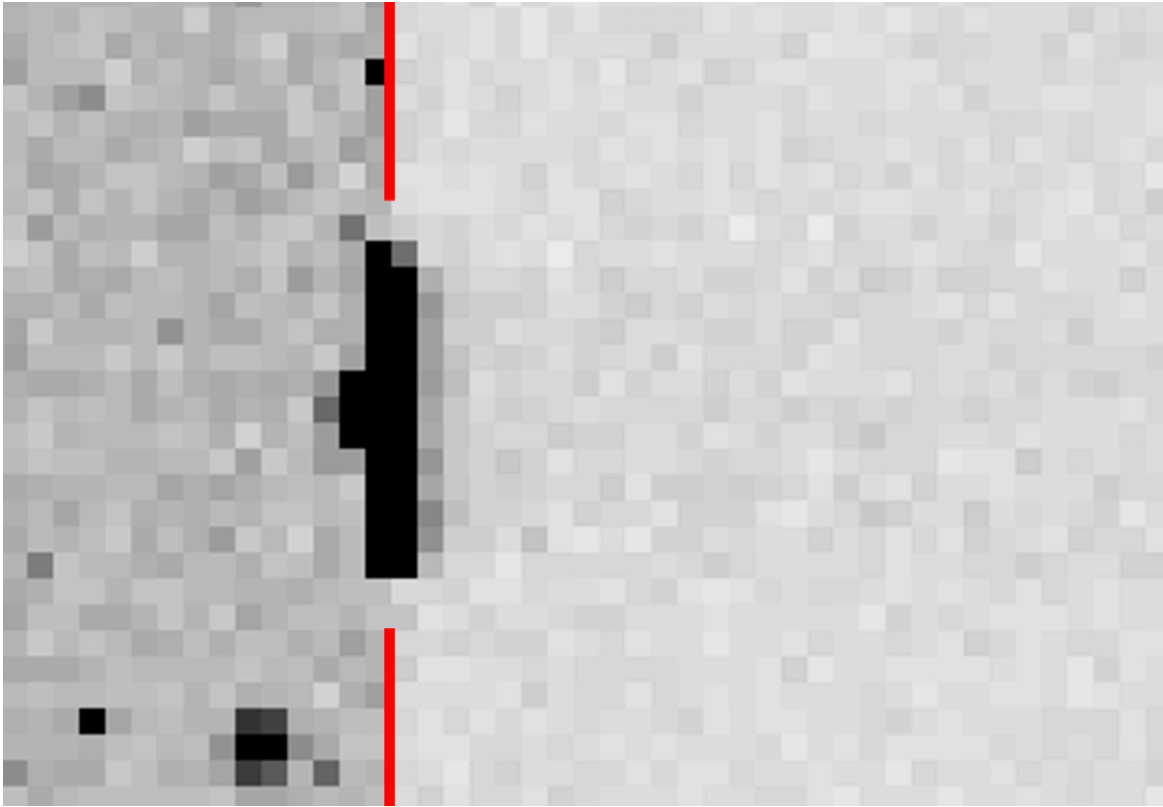

Figure 17: Image subsection of a saturated star that has a saturation bloom of about 10 pixels right at the amplifier boundary. The red line marks the boundary between the on-sky pixels (to the left) and the horizontal virtual overscan (to the right). This is a `_raz` format image, which has the pixels in the `_raw` images collected by amplifier and regularized in terms of readout direction.

## 5.4 Examining serial-trail wrapping

Now that we understand in principle how serial-CTE trails could wrap around to the next row, we searched for such cases in actual images to extend the analysis from **Figure 15** into the row above for the brightest stimulus pixels (saturated pixels between 50,000 e⁻ and 70,000 e⁻). In an effort to ensure sufficient stimulus to allow us to follow the trail as it fades, we identify all saturated stars with 3 to 15 pixels of blooming right at the amplifier boundary. There were 132 such stars right on the amplifier boundary in the 2023 full-frame GO images. One such example is shown in **Figure 17.**

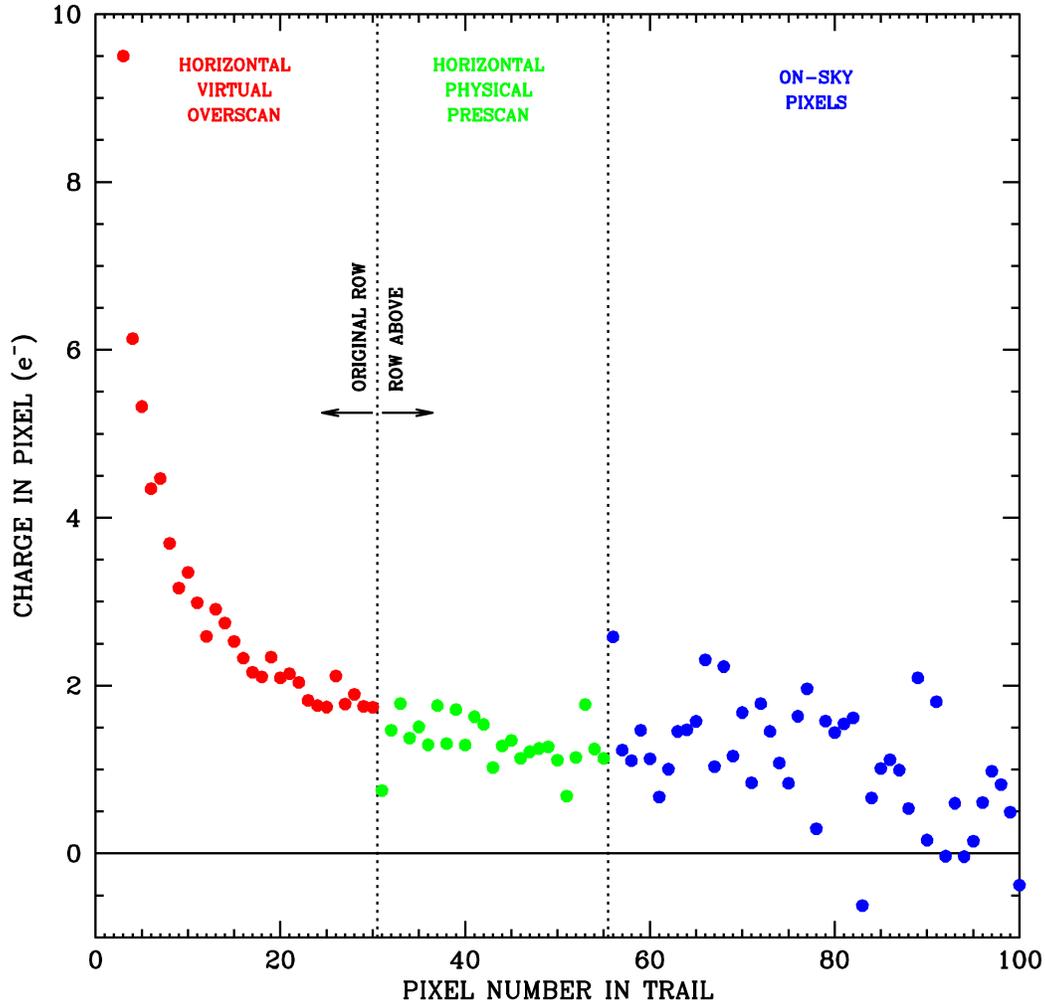

Figure 18: Trails behind pixels that are saturated in the last on-sky pixels (i=2048). The left region (**red**) shows the 30-pixel horizontal virtual overscan. The middle region (in **green**; pixels 31 through 56) show the 25 physical pre-scan pixels of the row above i.e., the row that followed the original row in the readout. Finally, the rightmost region (**blue**) shows the on-sky pixels of the row above.

**Figure 18** shows the trails behind pixels that are saturated at the edge of an amplifier boundary. The first 30 pixels in the trail (shown in red) come from the virtual horizontal overscan, which we studied in **Figures 14** through **16**. The next 25 pixels in the trail come from the physical pre-scan pixels (shown in green) in the row above the original row in the readout. Note that the trail continues at expected levels in the green points as if there was no change of row.

Trail pixels 56-100 (blue points) show that the same trend also continues seamlessly into the on-sky pixels, although with much more noise given the background (dark current, postflash, sky, and astrophysical sources). The pre/post-scan pixels contain essentially only readnoise. As such, the subtraction noise (not to mention the noise in the pixels themselves) is much lower than in the on-sky pixels.

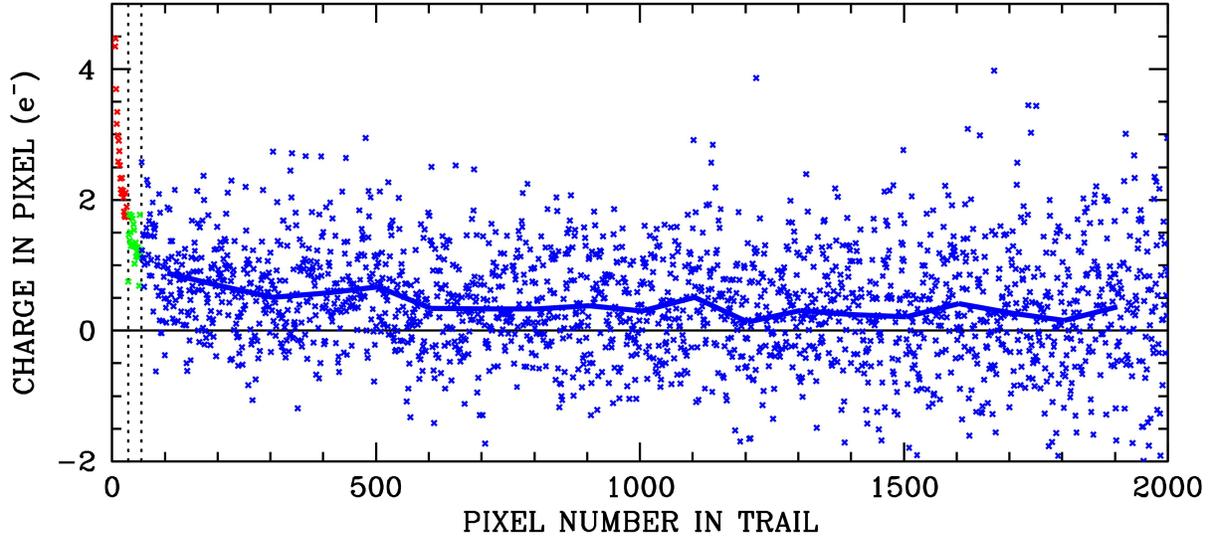

Figure 19: Continuation of Figure 18, the trails behind pixels that are saturated in the last on-sky pixels (i=2048).

**Figure 19** shows the full extent of the trailing from the saturated pixels into the next readout row. Even 2000 pixels out, the emptying traps deposit an average of about 0.25 e⁻ into each pixel. The charge between the 100$^{th}$ and the 2000$^{th}$ pixel of the trail totals about 760 e⁻, for stimulus pixels of ~60,000 e⁻. In other words, due to x-CTE, about 1% of the stimulus pixels' flux is deposited into far-out extended trails. Added to the 480 e⁻ in the first horizontal overscan pixel, 120 e⁻ in the rest of the horizontal overscan, and about 30 e⁻ in the 25 physical pre-scan pixels of the row above, the trails behind saturated pixels contain about 1390 e⁻, or roughly 2.5% of the stimulus-pixel flux.

## 5.5 Examining serial-trail wrapping using cosmic rays in CR-split long darks

Long low-level serial trails are clearly present, even though the noise in the on-sky pixels is such it is difficult to quantify the trails, particularly as they fade. For the extreme cosmic ray in **Figure 1**, the trail was in an unflashed bias image i.e. the lowest possible image background. Also, the phenomenon was first observed in a Quicklook image that had been binned to highlight faint features. In an effort to obtain improved signal-to-noise in the long serial-CTE trails, we examine CRs in the long dark images.

Standard long darks (900 sec) are taken with a postflash of 20 e⁻/pixel. A total of 1140 such darks were acquired during 2023. We focus here on the 682 dark frames taken in as back-to-back CR-split pairs. To isolate the contribution of CRs, we subtract one member of the pair from the other. This provides an exquisitely differential way to study the impact of the serial trails on the image.

We selected CRs that had at least 25,000 e⁻ in their maximum pixels and were located between columns i=1750 and i=2000 in each quadrant. For each CR, we recorded the baseline-subtracted[3] flux in each pixel of the trail, from $\Delta i = 0$ to $\Delta i = 2000$. The trail was followed into the prescan of the next row at the end of the horizontal overscan.

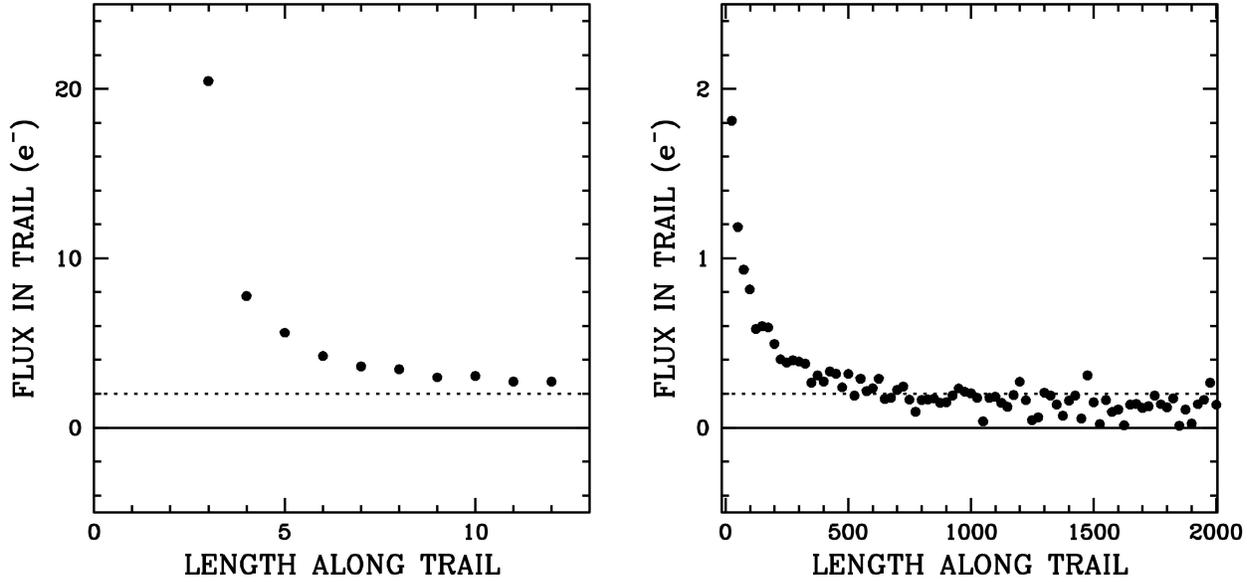

Figure 20: The x-CTE trail behind saturated (50000 e⁻ < p0 < 70000 e⁻) CRs occurring in on-sky columns i=1750 and i=2000 from the 682 paired darks. At left, the first ~dozen pixels trailing the saturated CRs (~2500 total). At right the full trails with results binned in groups of 25 trail pixels.

**Figure 20** shows the average trail behind the saturated CRs. The left panel shows the early few pixels in the trail (the first couple trail pixels are sometimes impacted by the CR itself). This is consistent both qualitatively and quantitatively with what we saw for the saturated stars at the amplifier edge (**Figure 19)**. The right panel shows the extended trail, binned into groups of 25 trail pixels. Again, the trend here is similar to that seen for the on-edge saturated stars in **Figure 19**, but much less noisy.

As is evident from the right panel of **Figure 15,** the trails behind different stimulus pixels have different scalings but similar shapes. **Figure 21** shows that the extended trails behind stimulus pixels that are about half-way to saturation have trails with about half the amplitude of trails following saturated pixels, consistent with **Figure 15**.

---

[3] Since darks can have slight fractional-pixel pedestal offsets from one another related to small bias drifts, we used a similar local background subtraction technique to that shown in **Figure 12** to isolate the contribution from the trail in each pixel.

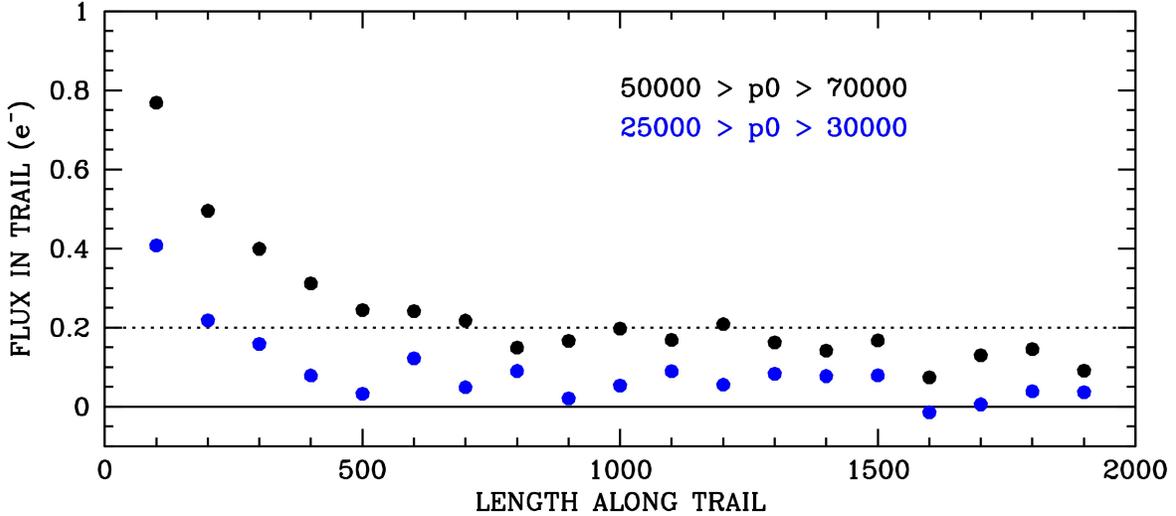

Figure 21: The same trail shown in Figure 20, but binned into 100-pixel wide bins. The trails behind pixels that are half-way and fully unsaturated are in blue and black, respectively.

## 6 Tabulating the serial trail model

After the analyses in the preceding sections, we now have enough information to construct a comprehensive model for the serial-CTE trails. There are two aspects to the trailing: (1) the total number of electron traps that a pixel packet of a given size will encounter if shuffled all the way across the 2048 pixels of the serial register, and (2) how these trapped electrons will be released.

With these two aspects of serial CTE, it should be straightforward to construct a simple software routine to model the serial trailing and remove it from images. Unlike parallel CTE, the impact of serial CTE on images is a small perturbation everywhere, so it should not require any algorithmic iterations and it should have a minimal impact on the image noise. The details of the serial correction algorithm will be slightly different from that for the parallel case, since in the parallel case the rows are all independent. In the serial case, one row can affect the row above it, via the wrapping effect. Thus, the x-CTE correction likely cannot be parallelized in the same manner as that for y-CTE. However, the x-CTE correction should intrinsically run much faster since there are fewer traps and there will be no iterations.

**Table 3** consolidates the trail information from **Figures 8** and **14** (recall that **Figure 14** has roughly two electrons removed, since there was an image background of 20 electrons). **Table 4** consolidates the information from **Figures 20** and **21**. These tabulations will form the basis for our pixel-based model.

Table 3: Distillation from Figures 8 and 14, listing the number of electrons needed in a stimulus pixel to generate a given number of electrons in the first serial trail pixel for each of the four amplifiers. The epoch corresponds to mid-2023, when WFC3/UVIS had been in orbit for 14 years.

| Electrons in first serial-CTE pixel | Size of Electron Cloud Needed | | | |
|---|---|---|---|---|
| | Amp A | Amp B | Amp C | Amp D |
| 1 | 7 | 7 | 7 | 7 |
| 2 | 48 | 48 | 48 | 48 |
| 3 | 73 | 73 | 73 | 73 |
| 4 | 112 | 165 | 135 | 95 |
| 5 | 150 | 250 | 200 | 110 |
| 6 | 200 | 350 | 275 | 125 |
| 7 | 250 | 550 | 350 | 150 |
| 8 | 300 | 700 | 375 | 200 |
| 9 | 375 | 900 | 400 | 250 |
| 10 | 400 | 1000 | 600 | 300 |
| 11 | 600 | 1100 | 700 | 350 |
| 12 | 700 | 1250 | 800 | 400 |
| 14 | 950 | 1500 | 1000 | 450 |
| 17 | 1250 | 2000 | 1500 | 700 |
| 20 | 1600 | 2400 | 2000 | 1000 |
| 25 | 2100 | 3000 | 3000 | 1400 |
| 30 | 2800 | 3700 | 3800 | 2100 |
| 40 | 4000 | 5500 | 6000 | 3500 |
| 50 | 7000 | 7500 | 8000 | 5500 |
| 75 | 12000 | 12000 | 15000 | 11000 |
| 100 | 15500 | 15500 | 17500 | 16000 |
| 125 | 17500 | 18000 | 20000 | 18000 |
| 150 | 20000 | 21000 | 25000 | 21000 |
| 200 | 24000 | 25000 | 27000 | 25000 |
| 300 | 32500 | 36000 | 37500 | 37000 |
| 400 | 42500 | 46500 | 47000 | 52000 |
| 500 | 56000 | 59000 | 58000 | 67500 |
| 600 | 67500 | 65000 | 65000 | 85000 |

**Table 4: Observed trail for ~60,000 e⁻ stimulus along with the implied cumulative distribution and fractional distribution (from Figures 20 and 21).**

| Trail Pixel | Observed (e⁻) | Cumulative (e⁻) | Fraction |
|---|---|---|---|
| 1 | 486 | 486 | 0.439700 |
| 2 | 31 | 517 | 0.028040 |
| 3 | 9.5 | 527 | 0.008594 |
| 4 | 6.1 | 533 | 0.005518 |
| 5 | 5.3 | 538 | 0.004795 |
| 6 | 4.6 | 543 | 0.004161 |
| 7 | 4.2 | 547 | 0.003800 |
| 8 | 3.8 | 551 | 0.003438 |
| 10 | 3.3 | 554 | 0.002985 |
| 15 | 2.5 | 571 | 0.002262 |
| 20 | 2.0 | 582 | 0.001809 |
| 30 | 1.75 | 601 | 0.001583 |
| 40 | 1.50 | 617 | 0.001357 |
| 50 | 1.20 | 630 | 0.001086 |
| 75 | 0.95 | 657 | 0.0008594 |
| 100 | 0.80 | 679 | 0.0007237 |
| 150 | 0.60 | 714 | 0.0005428 |
| 200 | 0.50 | 742 | 0.0004523 |
| 300 | 0.40 | 786 | 0.0003628 |
| 400 | 0.30 | 821 | 0.0002714 |
| 500 | 0.25 | 849 | 0.0002260 |
| 750 | 0.20 | 905 | 0.0001809 |
| 1000 | 0.175 | 952 | 0.0001583 |
| 1500 | 0.150 | 1033 | 0.0001357 |
| 2000 | 0.100 | 1096 | 0.0000905 |
| 2103 | 0.090 | 1105.4 | 0.0000811 |

# 7 A pixel-based correction model for serial CTE

The parameters in **Tables 3** and **4** are analogous to the parameters for y-CTE that have been used in the pixel-based correction software for years. We now employ them to construct a correction routine for the serial (x-)CTE.

## 7.1 Formulating the correction routine

The routine starts with an empty serial register, modelled as being 4206 pixels long to account for all of the "virtual" pixels that get shifted into the register during the 2103 pixels[4] that are read out.

Before reading out each row, we first find the maximum pixel value in that row. This will tell us how many traps we need to consider. We then cycle loop through the traps from $N_{MAX}$ to 1. Each trap has a particular threshold pixel value $P_N$ that dictates which charge cloud(s) it will affect (from interpolation of **Table 3** for the particular amp being read out). An inner loop goes through the pixels from $i = 1$ to $i = 4206$. If pixel $i$ has more than $P_N$ electrons, then $f$ electrons are removed from the pixel, where $f = i/2073$, and the trail counter for the trap $T$ is set to 0. With each subsequent pixel, the trail counter increases by one and we release $r$ electrons into the pixel, where $r = f \times R[T] / R[1]$, where $R[T]$ comes from interpolation of the fourth column of **Table 4**. If another pixel greater than $P_N$ is encountered, then the trap is flushed out[5], and the trap is filled again with a value of $f$ that is appropriate for the current column number, $i$. This process continues until all 4206 pixels are processed. Note that for the post-overscan pixels — $i > 2103$ — the value of $f$ is constrained to never be greater than 1.0, since even virtual pixels cannot be shifted more than 2103 times).

After the trapping and releasing has been modeled for all $N_{MAX}$ traps in a given row, the first 2103 pixels in the serial register are then inserted into a new copy of the 2103 × 2070 quadrant of the image. The next 2103 pixels in the serial register now correspond to the state of the serial register before the next image-pixel row is shifted in. This way, the serial register no longer starts out empty. Rather, it contains the released charge from the prior row's emptying traps (see **Figure 16**). Since traps have trails of 2103 pixels in length, there is always some fractional number of electrons in each serial-register pixel before the next image row is parallel-shuffled in from above. The procedure is repeated for all 2070 rows of the quadrant.

Once the entire quadrant has been read out this way, the new 2103 × 2070 image (**N**) corresponds to a forward application of the serial-CTE model — it represents what the observed image (**O**) would look like if it were serially read out one more time. The final step is to use this image to undo the trail effect.

---

[4] 2103 = 25 physical horizontal pre-scan, 2048 image pixels, and 30 virtual horizontal overscan.
[5] $\sum_{t}^{2103} r_t$ is released.

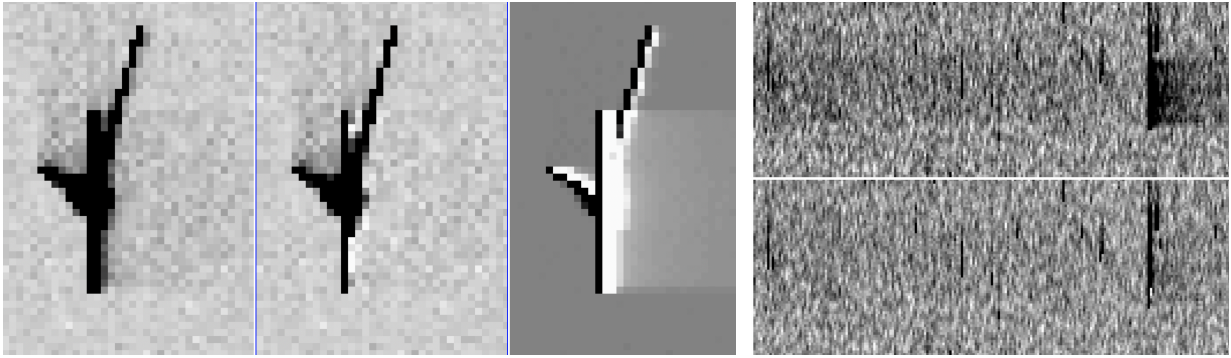

**Figure 22:** (Left) Close up of the massive CR bias event from Figure 1: the original, corrected, and difference. (Right) Block-averaged image of the original (top) image and corrected (bottom) image, showing all 2103 pixels of the amplifier from the pre-scan to the post-scan.

Since the serial-readout operation is such a mild perturbation, we simply compute the difference image $D = N - O$, which corresponds to how the electrons were rearranged by the serial shuffling. We then construct the corrected image $C = O - D$. Image $C$ should correspond to the image *before* serial readout. Of course, the reshuffling experienced by image $C$ will not be exactly the same reshuffling as was seen in image $O$, but it will be very close, and there should be no need for the iteration that is necessary with the y-CTE correction

The x-CTE correction is formulated as a FORTRAN subroutine named `sub_raz2rzx_wfc3uv`, taking as input a raz-format image and the date of observation in fractional years since 2000, and returning the x-CTE-corrected version of the same image. **Section 8** describes how this routine can be accessed.

## 7.2 Qualitatively evaluating the pixel-based x-CTE correction

The first test was to apply the correction to the extreme event that initially prompted this analysis (Figure 1). **Figure 22**, shows the original image, the corrected image, and the difference image. In the difference image, dark corresponds to flux added to pixels and white refers to flux subtracted from pixels (i.e., flux returned to the bright pixels and removed from the trailed pixels). The correction does a reasonable job removing the initial parts of the trail to the right of the CR event. The CR event is clearly not a delta function — neither in *x* nor in *y*, so it's difficult to know what a corrected version should look like. However, the trailing behind the secondary ~800 e$^-$ CR streak to the upper right from the main CR is very nicely removed, even in the very first trail pixel.

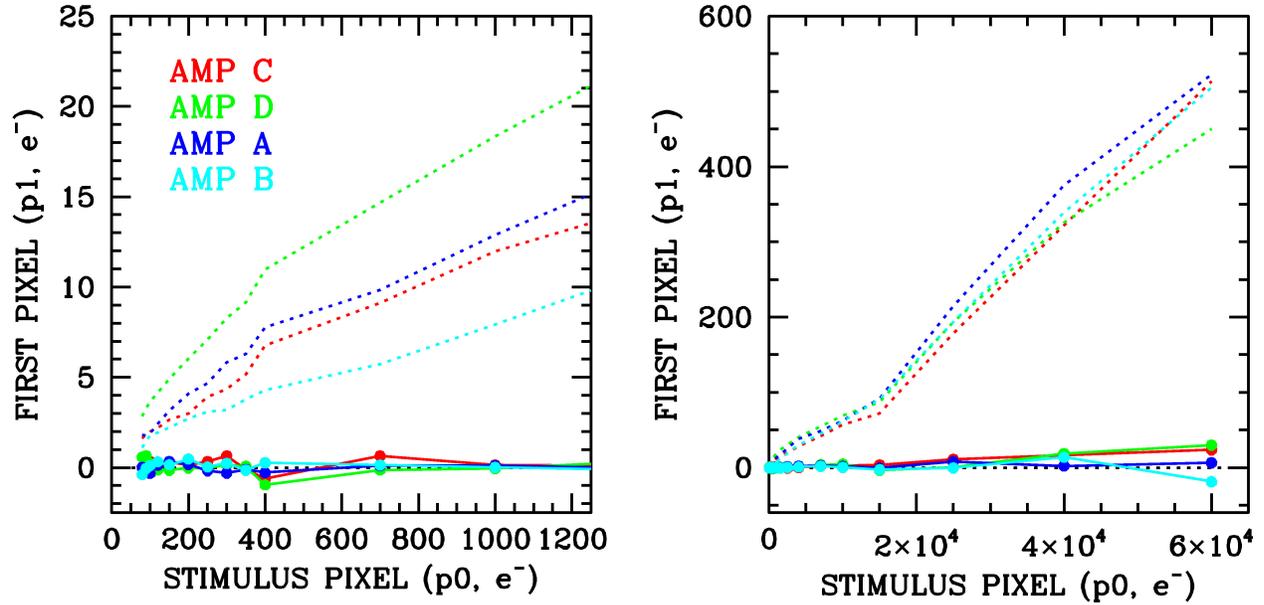

Figure 23: Same as Figure 14, except that here we show the uncorrected results as dotted curve and the results based on the pixel-based correction in solid.

## 7.4 Quantitatively evaluating the pixel-based correction

The next test for the pixel-based correction is the trails in the post-scan from **Section 5.2**. Recall that we examined the overscan pixels for which the pixel adjacent to the overscan (i=2073 in the \_raz images) was a maximum in the 11×11 box centered on that pixel. We recorded the upstream and downstream pixels and plotted the trend in the first trail pixel against the stimulus pixel as a function of amplifier in **Figure 14**.

In **Figure 23** we show that the pixel-based corrections do an excellent job removing the first pixel in the trail for stimulus pixels at all brightness levels in all four amplifier zones. The rest of the trail pixels are similarly well removed.

### 7.4.1 Impact on astrometry

While removing trails from unusual CR events in the bias frames or overscan provides a convenient demonstration of the correction's validity in extreme circumstances or non-sky pixels, it is of more interest to demonstrate how well the correction improves the on-sky pixels that contain astrophysical signal. In the original WFC3 study on serial CTE, Anderson (2014) examined the impact of serial CTE on astrometry. Here, we use the large set of data taken of the center of Omega Cen over the 15-year lifetime of WFC3/UVIS. There are ~200 images taken between 2009 and late 2023 of the cluster core at a variety of orientations with exposure times of ~60s through the popular F606W filter. These many images are intercompared to allow us to measure the impact of serial CTE on star measurements.

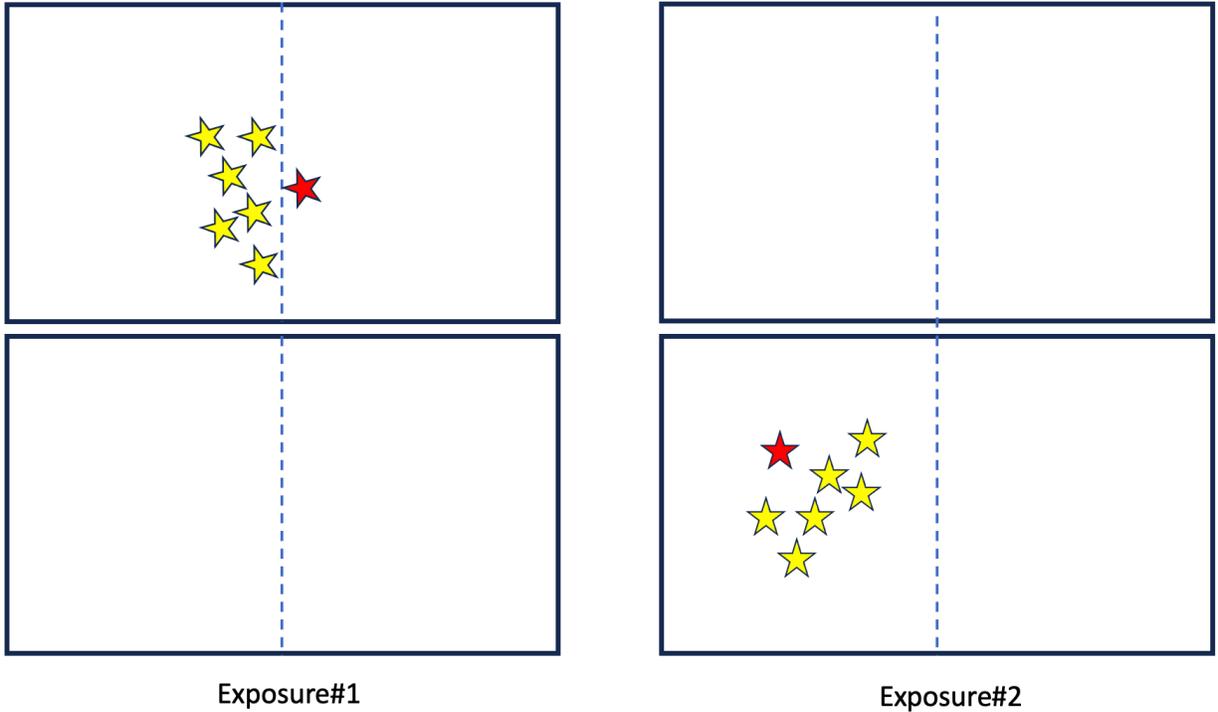

**Figure 24: Test used to evaluate the astrometric impact of x-CTE. In Exposure #1 there is a star just across the amplifier boundary (dotted line) from neighboring stars. In Exposure #2, all the stars are in the same amplifier region.**

**Figure 24** provides a schematic that demonstrates how we sense the astrometric impact of serial CTE. In Exposure#1, the yellow stars will all be serial-transferred to the left, but the red star will be serial-transferred to the right. If there is imperfect CTE in the serial direction, then the positions of the yellow stars in Exposure#1 will be shifted to the *right* towards the amplifier boundary, and the red star will be shifted to the *left* towards the amplifier boundary. Exposure#2 shows the *true* relative locations of the stars, since they are all on the same amplifier.

Measuring the shift across the amplifier boundary requires knowledge of the true position of the red star relative to the yellow stars. To determine this, we construct a 6-parameter linear transformation from Frame#2 to Frame#1 based on the yellow stars (correcting all positions for geometric distortion, of course). We then use that transformation to convert the Frame#2 position for the red star into a predicted position for that star in Frame#1. This position can then be compared against the observed position to determine the net astrometric shift. Since x-CTE (and y-CTE) can affect stars of different brightnesses differently, we must base the transformations on a limited brightness range of stars.

The target stars (the red stars) were chosen to all be within ~100 pixels of the amplifier boundary, and the comparison stars (the yellow stars) were all within ± 250 pixels of the red star, but on the opposite side of the boundary. Typically, there were between 25 and 100 comparison stars available for each target star. In practice, we use comparison stars that are within a factor of two of brightness of the target star and compare only images that are taken within 1.5 years of one another, to ensure that the random proper motions of stars (which have an amplitude of ~1 mas/year = 0.025 pixel/year for WFC3/UVIS at the center of Omega Cen) will not significantly blur out the astrometric signal we are seeking.

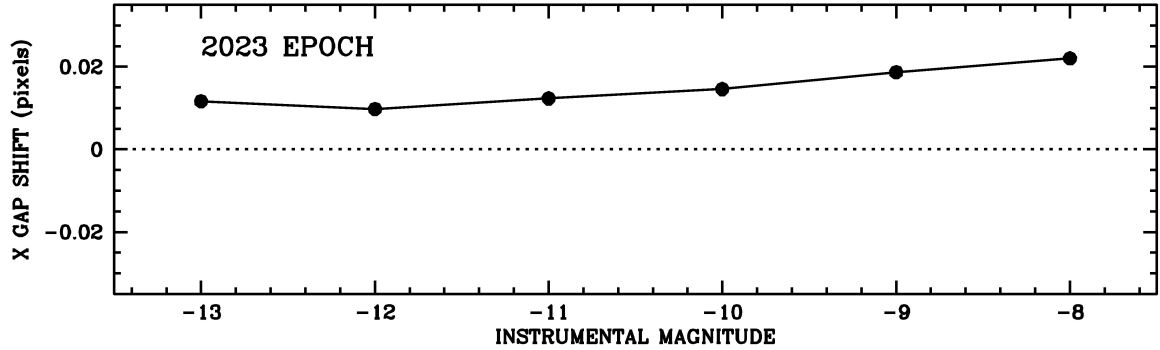

**Figure 25:** Shift across the amplifier boundary as a function of star brightness, as measured in images from 2023. Instrumental magnitude is m = −2.5 log$_{10}$(flux); saturation is at about m = −14.

**Figure 25** shows the x shift across the amplifier boundary for the most recent epoch i.e. the total shift between red and yellow stars. Referring to **Figure 24**, this shows that the bright yellow stars are shifted *right* by about 0.006 pixel and the red star is shifted *left* by 0.006 pixel, for a total shift of 0.012 pixel of the red relative to the yellow stars. Fainter stars experience somewhat larger shifts. These shifts are about 3 times those measured in the 2014 report (WFC3/ISR 2014-02), consistent with an x-CTE effect as WFC3/UVIS has now been in space for three times longer.

*These shifts may appear miniscule, but S/N ≥ 100 stars can be measured to about 0.01 pixel in a single exposure, so this is a significant effect that should be taken into account when high-precision astrometry is the goal.* **Figure 26** shows that the effect is linear with time, again consistent with the x-CTE-related interpretation.

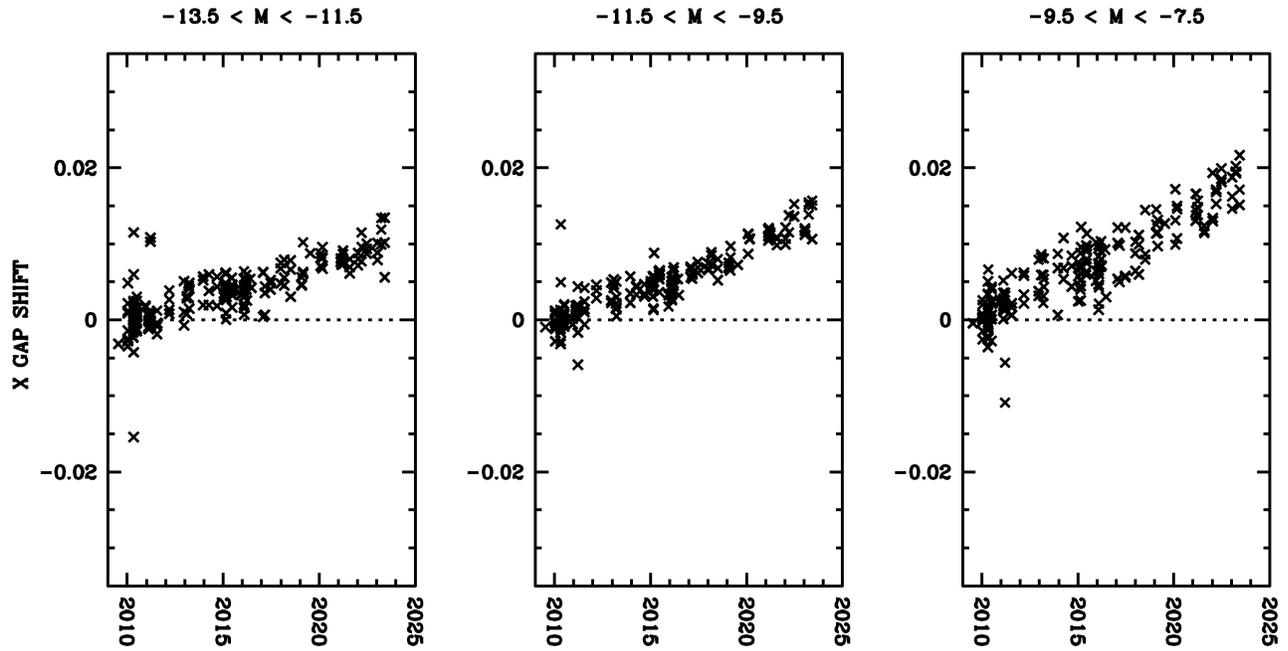

Figure 26: The amplifier-boundary shifts as a function of time for three bins of source brightness. The star-shifts for each of the ~200 exposures have been averaged together into a single point.

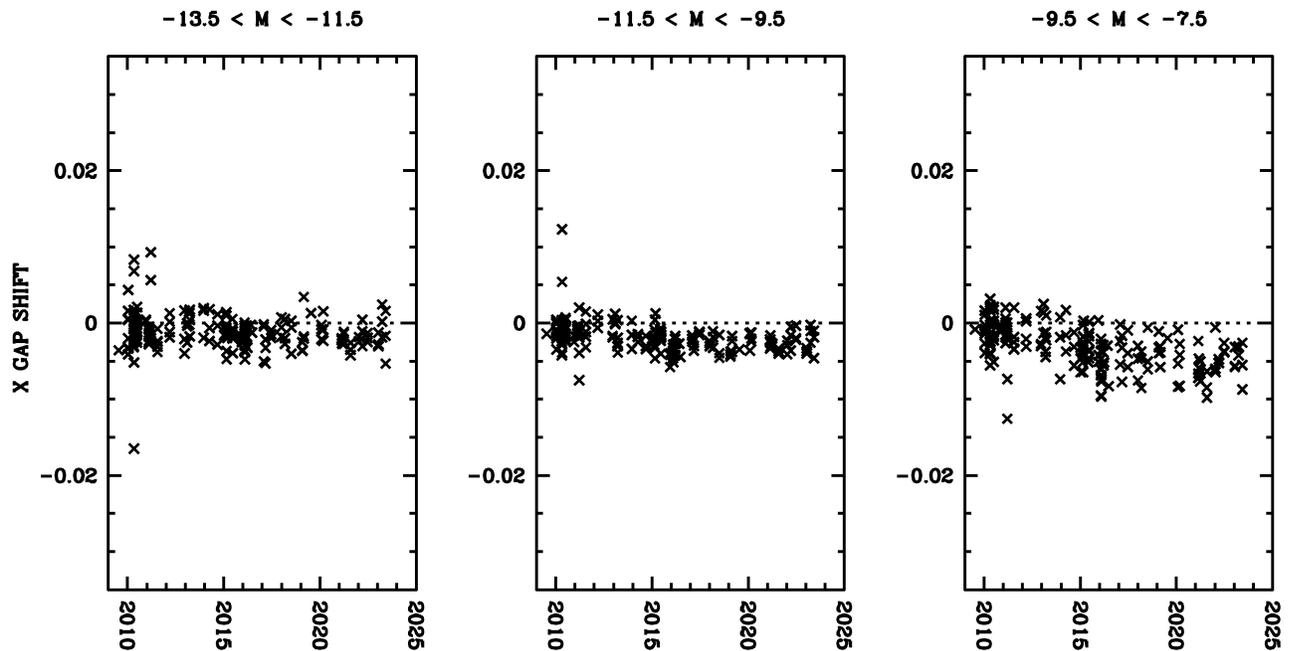

Figure 27: Same content as for Figure 26 though with measurements made on images corrected for serial-CTE using the pixel-based correction.

Figure 27 shows the data in Figure 26 post-correction. The trends are successfully removed for the brighter stars. However, the fainter stars (S/N ~ 50) appear to be over-corrected by perhaps 25%. It is worth noting that the net residual shift could be an indication that the distortion correction across the boundary may not be perfect. This would not be surprising given that the distortion correction in use is based on images taken between 2009 and 2011 (Bellini et. al. 2021), and there were likely small x-CTE trends at that time that were imprinted on the solution.

### 7.4.2 Impact on photometry

The maximum impact of imperfect CTE occurs at the amplifier boundary, i.e., the maximum distance from the amplifier. The astrometric signature is easy to see since the effect is in opposite directions on the two sides of the boundary. The photometric impact is more difficult to detect, since the impact is the same on both sides of the boundary.

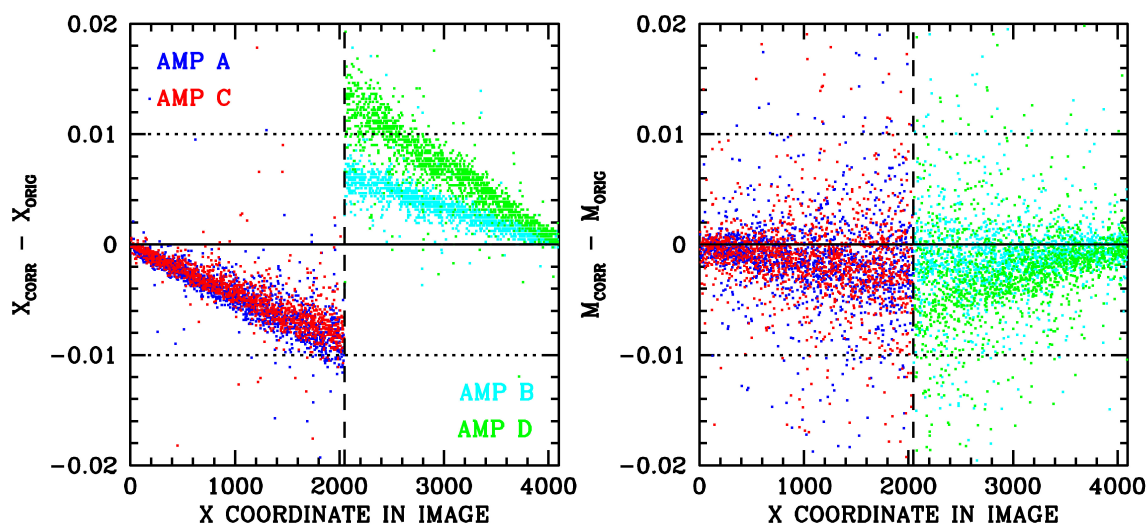

Figure 28: The impact of the x-CTE correction on astrometry (left) and photometry (right) for stars with S/N ~ 100 for images taken in 2023.

Even though it is hard to see the photometric impact of x-CTE directly in on-sky image comparisons, we can examine how big the pixel-based algorithm says the impact should be. In Figure 28 we compare the original positions and fluxes for stars with positions and fluxes measured for the same stars in the images after the x-CTE pixel-based correction has been applied. The images were from 2023 and the stars shown have S/N ~ 100. The results are separated by amplifier to show that the different trends seen in Figure 14 and recorded in Table 3 do have an impact on the corrections. The typical astrometric correction is about 0.0075 pixel, and the typical photometric correction is about 0.003 magnitude. The reason the photometric correction is so much smaller is that even though the serial trail goes out almost 2000 pixels, part of the trail is extremely sharp: about 50% of the captured flux is released in the first pixel of the trail. As such, more than half of the trail is captured within the 5×5-pixel aperture used in this analysis. By contrast, even though the y-CTE trail is only 100 pixels in length, only 20% is emitted into the first pixel. Also, stars have multiple bright pixels, and downstream pixels can provide some shielding for upstream pixels. For both of these reasons, even though the traps affect ~1% of the flux, the photometric impact is considerably less than 1%.

# 8. Applying the x-CTE correction

Most images are not significantly impacted by the x-CTE effect, however HST users that require high-precision astrometry could benefit from this correction. At the very least, they can quantify its impact on their science. The pixel-based serial CTE correction is not yet available in the automated calwf3 pipeline. Until then, it is available as a standalone correction via three routes:

(1) The pixel-based x-CTE correction can be run on the `_raw.fits` images to construct serial-CTE-corrected versions of the raw images. These images can then be run through the standard calwf3 pipeline. Note that to obtain a pixel-based correction for both serial and parallel CTE, the serial correction should be done *first*, since corrections should ideally be made in reverse order of the effects they are correcting. In this case, the serial-CTE blurring happened after the parallel-CTE blurring had already taken place, so it is most appropriate to correct serial-CTE before the parallel-CTE-correction pipeline is run. To run the serial correction on `_raw` images, first download the FORTRAN routine `raw2xcte_wfc3uv` from the WFC3 homepage[6] and compile it with `g77`. To run it, simply provide one (or more) WFC3/UVIS full-frame `*q_raw.fits` image names on the command line. The routine will generate a `*x_raw.fits` file that contains the corrected image that can then be processed through calwf3 pipeline.

(2) The pixel-based x-CTE correction can also be applied by using the FORTRAN routine `flt2xcte_wfc3uv`, which is also available from the WFC3 homepage. This routine will operate on a standard FLC or FLT image and will generate a new image (`*x_flt` or `*x_flc`) that has been pixel-corrected for serial-CTE. As pointed out in (1), the serial-CTE correction ideally should be performed before the parallel-CTE correction, but since the serial-CTE correction is small it is not critical and one can run the x-CTE correction on the y-CTE corrected `_flc` images. The user can then drizzle or run photometry routines (such as `hst1pass`) on the resulting image.

(3) The `hst1pass` software (Anderson, 2022), version 2024.05.29_v1g, now has the ability to run the pixel-based serial-CTE correction internally before it performs its PSF fitting for photometry and astrometry. The command-line instructions for `hst1pass` outline the usage. For more details, please see the report referenced above.

*Acknowledgements*

I thank Sylvia Baggett for noticing the strange event in the `if5ke4b3q` bias image through Quicklook and encouraging me to start this investigation. She also provided invaluable editorial feedback on early versions of this document. I am also grateful to Ben Kuhn for helping me ensure that the `_raw` files generated by `raw2xcte_wfc3uv` are compatible with the calwf3 pipeline.

---

[6] **https://www.stsci.edu/hst/instrumentation/wfc3/software-tools/cte-tools**